\documentclass{article}
\usepackage{amssymb}
\usepackage{amsmath}

\providecommand{\func}[1]{\mathop{\mathrm{#1}}\nolimits}

\begin{document}

\title{Quantum Field Theory on the Tangent Bundle}
\date{October 20, 2026}
\author{Joachim Herrmann\thanks{%
E-mail: \texttt{herrmannjoachim45@gmail.com}; ORCID: \texttt{%
0000-0002-2395-0135}.} \\
Formerly at Max Born Institute, now retired\\
Berlin, Germany}
\maketitle

\begin{abstract}
We develop the foundations of quantum field theory on the tangent bundle
(TB) in which internal gauge structures arise from vertical degrees of
freedom within a common geometric framework. The construction starts from an
affine tangent-fiber symmetry with Lorentz and translational sectors. The
translational sector is related to the coframe and to the teleparallel
description of gravity, whereas a reduced spectral Lie-algebra derived from
the Lorentz sector governs non-gravitational internal gauge degrees of
freedom. The stabilizer of a fixed nonzero null vector in the tangent fibers
is the massless little-group-like algebra of the Euclidean group $e(2)$;
projective representations motivate its central extension $e^{c}(2)$. A
separate compact $SU(2)$ structure is obtained from the unitary symmetry of
a twofold multiplicity of equivalent vertical representations in the
left-chiral matter sector. Similarly, an assumed threefold quark
multiplicity admits an $SU(3)_{c}$ copy symmetry, which is gauged locally.
Compatibility of the color center with vertical rotations is implemented
through an additional global quotient of a threefold rotation cover.
Together with the selected chiral matter spectrum, this permits the observed
fractional quark charges and accommodates additional hypothetical gauge
bosons associated with the extra generators. We formulate matter- and
gauge-field quantization, BRST symmetry and the relevant anomaly constraints
on the tangent bundle, and establish that the projected color sector has the
standard local QCD structure, including its one-loop beta function and
asymptotic freedom. Finally, the translational sector is related to
teleparallel gravity, and a structural action for possible additional
couplings to Lorentz-sector fields is formulated. The framework places
selected structural elements of particle physics and teleparallel gravity in
a common geometric setting, while leaving a detailed study of phenomena
beyond the Standard Model for future investigation.
\end{abstract}

1. \textbf{Introduction}

The Standard Model (SM) of elementary particles provides an exceptionally
successful description of the fundamental physical interactions at presently
accessible energies. Nevertheless, it leaves several empirical and
conceptual questions open: it does not explain the origin of its gauge
group, the observed representations of the fermions, the replication of
families, the origin of neutrino masses and the hierarchy of the observed
lepton and quark masses, a candidate for dark matter and others. It also
does not incorporate gravity. These limitations support the view that the SM
is an effective theory requiring extension. Although many extensions have
been proposed, none has so far received experimental confirmation. This
motivates frameworks in which symmetries and interactions arise from a more
fundamental geometric or quantum structure rather than being introduced
phenomenologically.

The geometric formulation of gravity and the gauge principle are two central
organizing ideas of modern physics. General relativity relates gravitation
to spacetime geometry, whereas gauge theory requires invariance under local
transformations acting on internal degrees of freedom in spacetime. In the
1960s, the equivalence between gauge theories and the mathematics of fiber
bundles was recognized revealing a deep relation between geometry and the
gauge theories describing non-gravitational interactions \cite%
{Lu,Trau,Wu,Ati,Nak,Is}. In this formulation, gauge potentials are principal
connections, matter fields are sections of associated bundles, and gauge
transformations act along the fibers at fixed spacetime points. The
fiber-bundle framework explains the geometric meaning of gauge fields, but
by itself it does not determine which fiber bundle, structure group, or
internal representations underlie the observed interactions.

In Refs.~\cite{I1,I2,I3} it was proposed to embed the SM gauge structure
into a deeper geometric framework based on the tangent bundle $TM$. A point
of $TM$ is locally specified by a spacetime coordinate $x$ and a tangent
vector $u\in T_{x}M$. Quantum fields may therefore depend not only on $x,$
but also on vertical variables parametrizing states in the tangent fiber.
The structure group of the tangent bundle is then related to that of an
affine frame bundle, with Lie-algebra $\mathfrak{g}_{\mathrm{TB}}=\mathfrak{%
so}(3,1)\ltimes \mathfrak{t}(3,1).$This algebra is isomorphic to the Poincar%
\'{e} algebra, however in the present construction it acts internally on
tangent or affine frames over a fixed base point; it is not assumed to
generate global spacetime isometries. We therefore use the qualifiers
TF-Lorentz and TF-Poincar\'{e} when confusion with spacetime transformations
could arise.

In Refs.~\cite{I1,I2,I3} it was proposed that the two sectors of the
structure group $\mathfrak{g}_{\mathrm{TB}}$ are assigned different physical
roles. The translational connection is identified with the coframe and
describes gravity in a form closely related to teleparallel gravity\cite{AL}%
. The TF-Lorentz-sector connection is used as the geometric source of
electroweak and additional internal gauge fields. This assignment differs
from Einstein--Cartan theory and from Poincar\'{e} gauge theory \cite%
{P1,P2,P3}, in which both the coframe and Lorentz connection are
gravitational variables. It is a defining hypothesis of the present TB
construction and must be tested through the consistency and phenomenology of
the resulting theory.

To obtain a tractable particle sector, we introduce a reduced vertical
Lie-algebra. A fixed nonzero null vector selected in the tangent fibers by
the vertical spectral problem has the stabilizer (little group) $e(2)$
within the TF-Lorentz group. Since particle-like vertical excitations are
taken to be bare-massless, this massless little-group-like reduction
provides a natural starting point. Projective representations lead to the
centrally extended algebra $e^{c}(2)$, whose Casimir has the form of a
two-dimensional Landau Hamiltonian. The lowest spectral level $n=0$ is
associated with a homogeneous vacuum condensate, whereas higher levels
describe particle-like excitations. Physical masses are expected to arise
from the interaction of higher Landau-level states with the vacuum
condensate; their calculation is outside the scope of this paper.

The compact weak-isospin structure has a different mathematical origin. When
the same irreducible vertical representation occurs twice in the left-chiral
matter sector, the two-dimensional copy space admits a unitary $U(2)$ action
commuting with $E^{c}(2)$. Its traceless part supplies the $SU(2)$ algebra
acting on electroweak doublets. Right-chiral fermions occupy separate
multiplicity-one sectors. Thus the reduced operator structure $SU(2)\otimes
E^{c}(2)$ is not a subgroup decomposition of $SO(3,1)$: the $E^{c}(2)$ part
acts on the irreducible vertical spectral space, whereas the $SU(2)$ part
acts on an equivalent-copy space.

The same representation-theoretic mechanism yields the color gauge group $%
SU(3)_{C}$. Three equivalent copies of a quark representation define a copy
space $\mathbb{C}_{c}^{3}$ with unitary symmetry $U(3)$. Its traceless part
is identified with the local color group $SU(3)_{c}$. The diagonal part $%
U(1)_{C}$ is not interpreted as a new gauge interaction independent of
electromagnetism. On a color triplet of fixed electric charge,
electromagnetic rotations give the same phase to all three color components.
We examine the compatibility of this phase with the center of the color
group, that is, the color transformations proportional to the identity
matrix. This extends the familiar global color--electromagnetic construction
to the centrally extended vertical group. Rotation periods and charge
normalization are explained explicitly in Section~3.

The paper develops the quantum-field-theoretic consequences of this
construction. We formulate the quantization of matter fields with continuous
vertical variables ($\xi _{1,}\xi _{2})$ and distinguish it from the
quantization of gauge connections. We then discuss BRST symmetry and anomaly
constraints. The diagonal anomaly conditions constrain possible relations
between hypercharge $Y$ and the central $E^{c}(2)$ character $\kappa $. The
charged $E^{c}(2)$ directions require a separate analysis and must not be
regarded as automatically covered by the Standard-Model anomaly conditions.
In the color sector, once the local $SU(3)_{c}$ connection and the quark
representations have been obtained, the BRST identities, renormalization
constants and one-loop beta function reduce to those of ordinary QCD. The
usual asymptotic freedom therefore follows.

Finally, we return to the translational sector. Its field strength is the
torsion, and the decomposition of the TF-Lorentz connection into
gravitational and internal contributions implies an extended torsion tensor
compared with the standard Teleparallel Gravity Theory~\cite%
{AL,t1,t2,t3,t4,g5}. This gives a possible additional coupling between
teleparallel gravity and non-gravitational gauge potentials. The proposed
geometrical coupling is specified at the structural level through the
extended gravity Lagrangian and the classical field equations; its
cosmological consequences, including any nonlinear response of the $n=0$
vacuum sector, require further dynamical calculation.

The paper is organized as follows. Section~2 introduces the affine
tangent-fiber geometry, the reduced vertical spectral Lie-algebra, the
multiplicity decomposition and the extended electroweak connection.
Section~3 constructs the $SU(3)_{c}$ copy symmetry and proposes a compatible
global structure with an explicitly normalized electromagnetic subgroup.
Section~4 treats quantization, BRST symmetry and anomaly constraints,
renormalization and the one-loop QCD beta function. Section~5 formulates a
possible structural extension of teleparallel gravity, and Section~6
contains the conclusions.

\bigskip \textbf{2. Geometric framework on the tangent bundle}

2.1 Structure group, generators and connections

Let $M$ be a four-dimensional spacetime manifold and 
\begin{equation}
TM=\bigsqcup_{x\in M}T_{x}M  \tag{2.1}  \label{2.1}
\end{equation}%
its tangent bundle. A local point of $TM$ is denoted by $(x^{\mu },u^{a})$,
where Greek indices refer to spacetime coordinates and Latin indices to a
local Lorentz frame. A frame $e_{a}=e_{a}{}^{\mu }\partial _{\mu }$ and its
dual coframe $e^{a}=e^{a}{}_{\mu }dx^{\mu }$ satisfy 
\begin{equation}
e^{a}{}_{\mu }e_{a}{}^{\nu }=\delta _{\mu }{}^{\nu },\qquad e^{a}{}_{\mu
}e_{b}{}^{\mu }=\delta ^{a}{}_{b}.  \tag{2.2}  \label{eq:old_2_2}
\end{equation}%
For $u,v\in T_{x}M$, the scalar product is 
\begin{equation}
g_{x}(u,v)=g_{\mu \nu }u^{\mu }v^{\nu }=\eta _{ab}u^{a}v^{b},\qquad \eta
_{ab}=\func{diag}(-1,1,1,1).  \tag{2.3}  \label{eq:old_2_3}
\end{equation}

Strictly, the linear frame bundle of a general four-manifold has structure
group $GL(4,\mathbb{R})$, reduced to $SO(3,1)$ after introduction of a
Lorentz metric and orientation. The semidirect product with translations
belongs to the corresponding affine-frame or Poincar\'{e}-gauge
construction. In the present paper we adopt this affine extension as the
tangent-fiber gauge structure, 
\begin{equation}
G_{\mathrm{TB}}=SO(3,1)\ltimes T(3,1).  \tag{2.4}  \label{2.4}
\end{equation}

On a principal bundle, the tangent space at a point $p$ splits into vertical
and horizontal subspaces once a connection has been chosen, 
\begin{equation}
T_{p}P=H_{p}P\oplus V_{p}P.  \tag{2.5}  \label{eq:old_2_5}
\end{equation}%
Vertical vectors generate transformations within a fiber, while horizontal
lifts define covariant comparison between neighboring fibers \cite{Is}. With
translation generators $P_{a}$ and TF-Lorentz generators $J_{ab}=-J_{ba}$,
the algebra is 
\begin{align}
\lbrack J_{ab},J_{cd}]& =i(\eta _{ad}J_{bc}+\eta _{bc}J_{ad}-\eta
_{ac}J_{bd}-\eta _{bd}J_{ac}),  \tag{2.6}  \label{eq:old_2_6} \\
\lbrack J_{ab},P_{c}]& =i(\eta _{bc}P_{a}-\eta _{ac}P_{b}),  \notag \\
\lbrack P_{a},P_{b}]& =0.  \notag
\end{align}%
The affine connection is written as 
\begin{equation}
\mathcal{A}_{\mu }=\theta ^{a}{}_{\mu }P_{a}+\frac{1}{2}\omega ^{ab}{}_{\mu
}J_{ab}.  \tag{2.7}  \label{eq:old_2_7}
\end{equation}

In gauge formulations of gravity, the coframe is related to the
translational gauge potential through the soldering structure. In the
convention adopted here we identify 
\begin{equation}
\theta ^{a}=e^{a}=e^{a}{}_{\mu }dx^{\mu }.  \tag{2.8}  \label{eq:old_2_8}
\end{equation}%
This identification is not a consequence of a vertical translation moving
the base point $x$; vertical gauge transformations act at fixed $x$. Rather,
it is the solder form that relates the internal translational space to the
tangent space of the base manifold \cite{Is}.

The translational and Lorentz field strengths are respectively 
\begin{equation}
T^{a}=d\theta ^{a}+\omega ^{a}{}_{b}\wedge \theta ^{b}  \tag{2.9}
\label{eq:old_2_9}
\end{equation}%
and 
\begin{equation}
R^{ab}=d\omega ^{ab}+\omega ^{a}{}_{c}\wedge \omega ^{cb}.  \tag{2.10}
\label{eq:old_2_10}
\end{equation}%
In components, 
\begin{align}
T^{a}{}_{\mu \nu }& =\partial _{\mu }\theta ^{a}{}_{\nu }-\partial _{\nu
}\theta ^{a}{}_{\mu }+\omega _{\mu }{}^{a}{}_{b}\theta ^{b}{}_{\nu }-\omega
_{\nu }{}^{a}{}_{b}\theta ^{b}{}_{\mu },  \tag{2.11}  \label{eq:old_2_11} \\
R^{ab}{}_{\mu \nu }& =\partial _{\mu }\omega ^{ab}{}_{\nu }-\partial _{\nu
}\omega ^{ab}{}_{\mu }+\omega _{\mu }{}^{a}{}_{c}\omega _{\nu
}{}^{cb}-\omega _{\nu }{}^{a}{}_{c}\omega _{\mu }{}^{cb}.  \tag{2.12}
\label{eq:old_2_12}
\end{align}%
For a Dirac spinor on a spacetime admitting a spin structure, the
geometrical spin connection acts as follows: 
\begin{equation}
D_{\mu }\Psi =\left( \partial _{\mu }+\frac{1}{2}\omega ^{ab}{}_{\mu }\Sigma
_{ab}\right) \Psi ,\qquad \Sigma _{ab}=\frac{1}{4}[\gamma _{a},\gamma _{b}].
\tag{2.13}  \label{eq:old_2_13}
\end{equation}

The pair $(e^{a}{}_{\mu },\omega ^{ab}{}_{\mu })$ defines a
metric-compatible Riemann--Cartan geometry when $\omega ^{ab}{}_{\mu
}=-\omega ^{ba}{}_{\mu }$ and $\gamma _{a}$ are the Dirac matrices. In
Einstein--Cartan theory torsion is algebraically related to the spin density
in the minimal action. Poincar\'{e} gauge theories allow more general
torsion- and curvature-dependent Lagrangians and may contain propagating
torsion \cite{P3}. The present construction uses a different physical
assignment: the translational sector supplies the gravitational coframe,
whereas selected components of the TF-Lorentz connection are projected onto
non-gravitational gauge fields. The consequences for teleparallel gravity
are developed in Section~5.

2.2 Reduced vertical groups and multiplicity spaces

The vertical spectral Lie-algebra is obtained from the linear TF-Lorentz
action on the tangent fibers. Introduce rotations and boosts 
\begin{equation}
L_{i}=\frac{1}{2}\epsilon _{ijk}J_{jk},\qquad K_{i}=J_{0i},  \tag{2.14}
\label{eq:old_2_14}
\end{equation}%
which obey 
\begin{align}
\lbrack L_{i},L_{j}]& =i\epsilon _{ijk}L_{k},  \tag{2.15}
\label{eq:old_2_15} \\
\lbrack L_{i},K_{j}]& =i\epsilon _{ijk}K_{k},  \notag \\
\lbrack K_{i},K_{j}]& =-i\epsilon _{ijk}L_{k}.  \notag
\end{align}%
Suppose that the vertical spectral problem selects a fixed nonzero null
vector in the tangent fibers 
\begin{equation}
\ell ^{a}=(1,0,0,1).  \tag{2.16}  \label{eq:old_2_16}
\end{equation}%
Its stabilizer in $SO(3,1)$ is isomorphic to the Euclidean group $E(2)$.
Defining 
\begin{equation}
T_{1}=K_{1}+L_{2},\qquad T_{2}=K_{2}-L_{1},\qquad T_{3}=L_{3},  \tag{2.17}
\label{eq:old_2_17}
\end{equation}%
one obtains 
\begin{equation}
\lbrack T_{1},T_{2}]=0,\qquad \lbrack T_{3},T_{1}]=iT_{2},\qquad \lbrack
T_{3},T_{2}]=-iT_{1}.  \tag{2.18}  \label{eq:old_2_18}
\end{equation}

The use of projective unitary representations motivates the centrally
extended algebra $e^{c}(2)$, 
\begin{equation}
\lbrack T_{1},T_{2}]=iE,\qquad \lbrack T_{3},T_{1}]=iT_{2},\qquad \lbrack
T_{3},T_{2}]=-iT_{1},\qquad \lbrack E,T_{a}]=0.  \tag{2.19}
\label{eq:old_2_19}
\end{equation}%
The generator $E$ is called central because it commutes with every
generator, as the last relation in (\ref{eq:old_2_19}) states. Its
exponential gives the additional phase rotation. The distinction is
important: $E(2)$ is a subgroup of the Lorentz group, whereas $E^{c}(2)$ is
a central extension used for the vertical quantum representation and is not
itself a subgroup of $SO(3,1)$.

The compact $SU(2)$ factor does not arise from the unused Lorentz
generators. If the same irreducible representation occurs several times
inside the full vertical Hilbert space of the Lie-group $E^{c}(2)$ it can be
decomposed into irreducible components according to 
\begin{equation}
\mathcal{H}=\bigoplus_{\lambda }\mathbb{C}^{m_{\lambda }}\otimes \mathcal{H}%
_{\lambda }.  \tag{2.20}  \label{eq:old_2_20}
\end{equation}%
Here $\mathcal{H}_{\lambda }$ denotes one elementary type of representation
of the symmetry group $E^{c}(2)$ and $m_{\lambda }$ is the multiplicity with
which this representation occurs. The group $E^{c}(2)$ acts only on \ the
irreducible factor $\mathcal{H}_{\lambda }$ while it acts trivally on the
multiplicity factor $\mathbb{C}^{m_{\lambda }}.$ For a multiplicity $%
m_{\lambda }$ the multiplicity space has the dimension $m_{\lambda }.$ Its
natural compact symmetry is $U(m_{\lambda })$ with non-Abelian special
unitary part $SU(m_{\lambda })$ which acts on the multiplicity factor and
commutes with $E^{c}(2)$. A twofold multiplicity therefore supplies $U(2)$,
while a threefold multiplicity supplies $U(3)$. Their traceless parts are
the compact groups $SU(2)$ and $SU(3)$ used below. The commutant means the
set of operators that commute with the represented vertical symmetry. For
the mathematical treatment of multiplicity spaces and its emergent
symmetries see e.g.\cite{Hall,MS1,MS2} . This multiplicity must not be
confused with degeneracy inside a single $E^{c}(2)$ representation. The
latter is spanned by the vertical functions 
\begin{equation}
\mathcal{H}_{E^{c}(2)}=\func{span}\{h_{nm\kappa }(\xi )\},  \tag{2.21}
\label{eq:old_2_21}
\end{equation}%
whereas a doublet block has the factorized form 
\begin{equation}
\mathcal{H}_{D}=\mathbb{C}^{2}\otimes \mathcal{H}_{\lambda _{D}}.  \tag{2.22}
\label{eq:old_2_22}
\end{equation}%
The associated field carries a copy index $\alpha =1,2$, 
\begin{equation}
\psi _{\alpha }(x,\xi )=\sum_{n,m,\kappa }\psi _{\alpha nm\kappa
}(x)h_{nm\kappa }(\xi ).  \tag{2.23}  \label{eq:old_2_23a}
\end{equation}%
The $E^{c}(2)$ generators act on $h_{nm\kappa }$, while the multiplicity
group acts on $\alpha $. The two actions therefore commute.

In the electroweak interpretation, a left-chiral doublet and the associated
right-chiral singlet blocks have the general structure 
\begin{equation}
\mathcal{H}_{\mathrm{ferm}}=\left( \mathbb{C}_{L}^{2}\otimes \mathcal{H}%
_{\lambda _{L}}\right) \oplus \bigoplus_{r}\left( \mathbb{C}_{Rr}^{1}\otimes 
\mathcal{H}_{\lambda _{Rr}}\right) .  \tag{2.24}  \label{eq:old_2_23}
\end{equation}%
When the right-chiral blocks are inequivalent, the commutant of the vertical
representation on these blocks contains 
\begin{equation}
U(2)_{L}\times \prod_{r}U(1)_{Rr}.  \tag{2.25}  \label{eq:old_2_24}
\end{equation}%
The $SU(2)$ subgroup of $U(2)_{L}$ mixes the two equivalent left-chiral
copies and is identified with weak isospin. The diagonal generator of $%
U(2)_{L}$ is proportional to $\mathbf{1}_{2}$ and gives a common phase to
both members of the left-chiral doublet. It does not act on a separate
one-dimensional space and therefore cannot itself be identified with a
right-chiral multiplicity-one sector. Each right-chiral singlet belongs to a
separate block and has its own one-dimensional phase symmetry.

On these blocks a weak transformation acts as 
\begin{equation}
U_{\mathrm{weak}}=U_L\oplus\bigoplus_r\mathbf{1}_{Rr},\qquad U_L\in SU(2)_L.
\tag{2.26}  \label{eq:old_2_25}
\end{equation}

The weak-hypercharge transformation $U(1)_Y$, generated here by $T_3$, can
correlate the phases of the different blocks. Its representation has the
form 
\begin{equation}
Y=y_{L}\mathbf{1}_{2}\oplus \bigoplus_{r}y_{Rr},  \tag{2.27}
\label{eq:old_2_26}
\end{equation}%
and hence 
\begin{equation}
\exp (i\alpha Y)=e^{i\alpha y_{L}}\mathbf{1}_{2}\oplus
\bigoplus_{r}e^{i\alpha y_{Rr}}.  \tag{2.28}  \label{eq:old_2_27}
\end{equation}%
Accordingly, the diagonal $U(1)$ contained in $U(2)_{L}$ can be interpreted
as the restriction of the hypercharge transformation $\exp(i\alpha Y)$ to
the left-chiral doublet. This is not yet the electromagnetic transformation: 
$Q=J_3+Y/2$ gives different charges to the two doublet components after
symmetry breaking. It is neither the right-chiral singlet sector nor, by
itself, a new physical gauge interaction. In the present TB model the local
hypercharge direction is supplied by $T_{3}$ in the $E^{c}(2)$ sector;
consistency requires that its eigenvalues reproduce the blockwise action
displayed above.

Thus a twofold multiplicity of the irreducible representation of $E^{c}(2)$
naturally gives rise to an $SU(2)$ -type internal Lie algebra acting on the
multiplicity degrees of freedom. This point is especially useful when the
original vertical algebra is non-compact or non-semisimple, because the
compact $SU(2)$ structure can arise from the commutant rather than from the
original algebra itself.

Based on this construction a convenient low-energy basis is therefore 
\begin{align}
|\chi _{1}\rangle & =|D,+\rangle \otimes |n_{D},m_{D},\kappa _{D}\rangle , 
\tag{2.29}  \label{eq:old_2_28} \\
|\chi _{2}\rangle & =|D,-\rangle \otimes |n_{D},m_{D},\kappa _{D}\rangle , 
\notag \\
|\chi _{Rr}\rangle & =|S_{r}\rangle \otimes |n_{Rr},m_{Rr},\kappa
_{Rr}\rangle .  \notag
\end{align}%
The equality of the vertical labels within the doublet expresses the
equivalence required by the multiplicity construction. The labels of a
right-chiral singlet need not coincide with those of the doublet.

2.3 Generators and vertical spectrum of $E^{c}(2)$

Let $(\xi _{1},\xi _{2},\beta )$ be coordinates on the representation space.
A convenient differential realization for the generators on the group is 
\cite{r10,r11,I1}, 
\begin{align}
\mathbf{T}_{1}& =-i\left( \frac{\partial }{\partial \xi _{1}}+\frac{1}{2}\xi
_{2}\frac{\partial }{\partial \beta }\right) ,  \tag{2.30}
\label{eq:old_2_29} \\
\mathbf{T}_{2}& =-i\left( \frac{\partial }{\partial \xi _{2}}-\frac{1}{2}\xi
_{1}\frac{\partial }{\partial \beta }\right) ,  \notag \\
\mathbf{T}_{3}^{(0)}& =-i\left( \xi _{1}\frac{\partial }{\partial \xi _{2}}%
-\xi _{2}\frac{\partial }{\partial \xi _{1}}\right) ,  \notag \\
\mathbf{E}& =-i\frac{\partial }{\partial \beta }.  \notag
\end{align}%
The quadratic invariant is 
\begin{equation}
\mathbf{\Delta }_{0}=\mathbf{T}_{1}^{2}+\mathbf{T}_{2}^{2}+2\mathbf{T}%
_{3}^{(0)}E.  \tag{2.31}  \label{eq:old_2_30}
\end{equation}%
In polar coordinates $\xi _{1}=\xi \cos \varphi $ and $\xi _{2}=\xi \sin
\varphi $, simultaneous eigenfunctions of $\mathbf{E}$ and the reference
generator $\mathbf{T}_{3}^{(0)}$ can be written as

\begin{equation}
h_{nm\kappa }(\xi ,\beta ,\varphi )=e^{i\kappa \beta }e^{im\varphi
}g_{nm\kappa }(\xi ).  \tag{2.32}  \label{eq:old_2_31}
\end{equation}%
The radial equation is 
\begin{equation}
\left[ -\frac{1}{\xi }\frac{d}{d\xi }\left( \xi \frac{d}{d\xi }\right) +%
\frac{m^{2}}{\xi ^{2}}+\frac{\kappa ^{2}\xi ^{2}}{4}+\kappa m\right]
g_{nm\kappa }(\xi )=\epsilon _{nm\kappa }g_{nm\kappa }(\xi ).  \tag{2.33}
\label{eq:old_2_32}
\end{equation}%
For $\kappa\ne0$, the normalizable solutions of this reference Casimir are

\begin{equation}
g_{nm\kappa }(\xi )=N_{nm\kappa }e^{-|\kappa |\xi ^{2}/4}\xi
^{|m|}L_{n}^{|m|}(|\kappa |\xi ^{2}/2),  \tag{2.34}  \label{eq:old_2_33}
\end{equation}%
where $L_{n}^{|m|}$ is an associated Laguerre polynomial. The eigenvalues
are 
\begin{equation}
\epsilon _{nm\kappa }=|\kappa |(2n+|m|+1)+\kappa m.  \tag{2.35}
\label{eq:old_2_34}
\end{equation}%
Here $n=0,1,2,\ldots $ and $m\in \mathbb{Z}$. When the central coordinate $%
\beta $ is compact with period $2\pi $, single-valuedness gives $\kappa \in 
\mathbb{Z}$. For separate radial normalization $\int_{0}^{\infty }\xi \,d\xi
\,|g_{nm\kappa }|^{2}=1$, 
\begin{equation}
N_{nm\kappa }=\left[ \frac{|\kappa |^{|m|+1}n!}{2^{|m|}(n+|m|)!}\right]
^{1/2}.  \tag{2.36}  \label{eq:old_2_35}
\end{equation}%
Angular and central factors $\varkappa $ are normalized separately with the
chosen measure. These formulas correct the radial equation, functions and
eigenvalues quoted in Ref.~\cite{I1}: they follow directly by squaring the
generators with their factors $1/2$ and adding $2\mathbf{T}_{3}^{(0)}\mathbf{%
E}$.\footnote{%
With the generator and Casimir definitions of Eqs. (26)--(27) in Ref. [I1],
the radial operator contains $\kappa ^{2}\xi ^{2}/4$ , and the Gaussian
factor $exp(-\mid \varkappa \mid \xi ^{2}/4).$ These expressions correct the
normalization of the radial solutions previously quoted in that reference.}
The scalar differential operator agrees, for positive $\kappa $, with the
spin-independent electron Landau Hamiltonian in symmetric gauge. For a
comparison we define that lengths are measured in units of the magnetic
lenght $l_{B}=(\hbar $ $/\mid e\mid B)^{1/2}$ in SI units.For $\varkappa
\succ 0,$ the coordinate identification $r/l_{B}=\mid \varkappa \mid
^{1/2}\xi $ gives $\Delta _{\varkappa }/\varkappa =2H_{Landau}/(\hbar \omega
_{c})$ where $\omega _{c}=\mid e\mid B/\mu ,$ $\mu $ denotes the elctron
mass. The Landau spectrum and quantum-Hall applications are discussed in
Refs.~\cite{LiWang,Quinn}. In particular, the electron charge convention
must be distinguished from the positive quantity $|e|B$. Shifted global
representations used for quarks are specified in Section~3.3; there $m$
remains the integer orbital index and need not equal the hypercharge $Y$.

For $\kappa =0$ the Landau-level-like structure disappears and the
representation reduces to the unextended $E(2)$ case. The multiplicity $SU(2)
$ has the usual generators $\mathbf{J}_{i}$, distinct from the Lorentz
rotations $\mathbf{L}_{i}$, and Casimir 
\begin{equation}
\Delta _{SU(2)}=\mathbf{J}_{1}^{2}+\mathbf{J}_{2}^{2}+\mathbf{J}%
_{3}^{2},\qquad \Delta _{SU(2)}|j,j_{3}\rangle =j(j+1)|j,j_{3}\rangle . 
\tag{2.37}  \label{eq:old_2_36}
\end{equation}

2.4 Gauge potentials and gauge fields in the extended electroweak sector

The reduced vertical construction contains two commuting algebraic actions:
the multiplicity algebra $su(2)$ with generators $\emph{J}_{i}$, and the
centrally extended spectral algebra $e^{c}(2)$ with generators $\{\mathbf{T}%
_{1},\mathbf{T}_{2},\mathbf{T}_{3},\mathbf{E}\}$. The gauge connection is
written as 
\begin{equation}
\mathcal{W}_{\mu }=g_{1}W_{\mu }^{i}\mathbf{J}_{i}+g_{2}B_{\mu }^{a}\mathbf{T%
}_{a}+g_{3}C_{\mu }\mathbf{E},\qquad a=1,2,3.  \tag{2.38}
\label{eq:old_2_37}
\end{equation}%
The fields $W_{\mu }^{i}$ are identified with the weak gauge potentials. The
field $B_{\mu }^{3}$, abbreviated below as $B_{\mu }$, supplies the neutral
hypercharge-like direction before symmetry breaking. The potentials $B_{\mu
}^{1}$ and $B_{\mu }^{2}$ belong to the additional translation-like
directions $\mathbf{T}_{1}$ and $\mathbf{T}_{2}$ of $E^{c}(2)$, and $C_{\mu }
$ is associated with the central generator $\mathbf{E}$. It is convenient to
define 
\begin{equation}
W_{\mu }^{\pm }=\frac{1}{\sqrt{2}}(W_{\mu }^{1}\mp iW_{\mu }^{2}),\qquad
B_{\mu }^{\pm }=\frac{1}{\sqrt{2}}(B_{\mu }^{1}\mp iB_{\mu }^{2}). 
\tag{2.39}  \label{eq:old_2_38}
\end{equation}

The $SU(2)$ curvature is the standard expression 
\begin{equation}
W^{i}{}_{\mu \nu }=\partial _{\mu }W_{\nu }^{i}-\partial _{\nu }W_{\mu
}^{i}+g_{1}\epsilon ^{i}{}_{jk}W_{\mu }^{j}W_{\nu }^{k}.  \tag{2.40}
\label{eq:old_2_39}
\end{equation}%
The $E^{c}(2)$ curvature must be calculated from its own structure
constants. With one common coupling $g_{2}$ and the commutator convention
stated above, a useful component form is 
\begin{align}
F^{1}{}_{\mu \nu }& =\partial _{\mu }B_{\nu }^{1}-\partial _{\nu }B_{\mu
}^{1}-g_{2}(B_{\mu }B_{\nu }^{2}-B_{\nu }B_{\mu }^{2}),  \tag{2.41}
\label{eq:old_2_40} \\
F^{2}{}_{\mu \nu }& =\partial _{\mu }B_{\nu }^{2}-\partial _{\nu }B_{\mu
}^{2}+g_{2}(B_{\mu }B_{\nu }^{1}-B_{\nu }B_{\mu }^{1}),  \notag \\
F^{3}{}_{\mu \nu }& =\partial _{\mu }B_{\nu }-\partial _{\nu }B_{\mu }, 
\notag \\
F^{E}{}_{\mu \nu }& =\partial _{\mu }C_{\nu }-\partial _{\nu }C_{\mu }+\frac{%
(g_{2})^{2}}{g_{3}}(B_{\mu }^{1}B_{\nu }^{2}-B_{\nu }^{1}B_{\mu }^{2}). 
\notag
\end{align}%
If different coupling constants are absorbed into the four connection
components, the coefficients in these equations change accordingly. The
central curvature $F^{E}{}_{\mu \nu }$ is therefore Abelian in its
derivative term but contains a bilinear contribution fixed by $[\mathbf{T}%
_{1},\mathbf{T}_{2}]=i\mathbf{E}$.

\bigskip

2.5 Invariant form, gauge Lagrangian and gauge transformations

Since $e^{c}(2)$ is non-semisimple, its Killing form is degenerate. A
Yang--Mills kinetic term must instead use a nondegenerate invariant
symmetric bilinear form $h_{AB}$ satisfying the gauge invariance condition 
\begin{equation}
h_{AD}f^{D}{}_{BC}+h_{BD}f^{D}{}_{AC}=0.  \tag{2.42}  \label{eq:old_2_41}
\end{equation}%
In the ordered basis $(T_{1},T_{2},T_{3},E)$, one convenient choice is \cite%
{Na,Tsey}

\bigskip\ 
\begin{equation}
(h_{AB})=%
\begin{pmatrix}
1 & 0 & 0 & 0 \\ 
0 & 1 & 0 & 0 \\ 
0 & 0 & k & 1 \\ 
0 & 0 & 1 & 0%
\end{pmatrix}%
,  \tag{2.43}  \label{eq:old_2_42}
\end{equation}%
where $k$ is arbitrary. The determinant is nonzero, although the Killing
form is degenerate. The neutral block has one positive and one negative
eigenvalue, so nondegeneracy alone does not establish a positive physical
state space. Covariant gauge fixing and Faddeev--Popov ghosts give a formal
quantization, as implemented in Ref.~\cite{I1}. The negative internal
kinetic direction also affects transverse vector modes and is not
automatically removed by ordinary gauge fixing. Positivity of the BRST
physical space therefore requires a further argument, such as a consistent
decoupling or restriction. Tseytlin discusses this possibility for the pure
gauge model, without establishing it for the matter-coupled construction
considered here \cite{Tsey}. The gauge-field Lagrangian can be written as 
\begin{equation}
\mathcal{L}_{\mathrm{gauge}}=-\frac{1}{4}W^{i}{}_{\mu \nu }W_{i}{}^{\mu \nu
}-\frac{1}{4}\widetilde h_{AB}F^{A}{}_{\mu \nu }F^{B\mu \nu }.  \tag{2.44}
\label{eq:old_2_43}
\end{equation}%
Here the algebra-valued curvature is $\mathcal{F}_{\mu\nu}=g_2F^a_{\mu%
\nu}T_a+g_3F^E_{\mu\nu}E$. The kinetic form in these component fields is $%
\widetilde h_{AB}=g_Ag_Bh_{AB}$, with $g_a=g_2$ and $g_E=g_3$. Consequently, 
\begin{equation}
\mathcal{L}_{E^c(2)}=-\frac14\left[ g_2^2\left(F^1_{\mu\nu}F^{1\mu\nu}+F^2_{%
\mu\nu}F^{2\mu\nu} +kF^3_{\mu\nu}F^{3\mu\nu}\right)
+2g_2g_3F^3_{\mu\nu}F^{E\mu\nu}\right].  \tag{2.45}  \label{eq:old_2_44}
\end{equation}
A common overall coupling factor can be absorbed into the kinetic
normalization. The invariant form and the field normalization must always be
used together.

For an algebra-valued parameter $\Theta =\theta ^{A}T_{A}$, the
infinitesimal gauge transformation is 
\begin{equation}
\delta \mathcal{W}_{\mu }=D_{\mu }\Theta =\partial _{\mu }\Theta -i[\mathcal{%
W}_{\mu },\Theta ],  \tag{2.46}  \label{eq:old_2_45}
\end{equation}%
In components with a common coupling, 
\begin{equation}
\delta A_{\mu }^{A}=\partial _{\mu }\theta ^{A}+g_{2}f^{A}{}_{BC}A_{\mu
}^{B}\theta ^{C}.  \tag{2.47}  \label{eq:old_2_46}
\end{equation}%
In particular, the central potential receives the term 
\begin{equation}
\delta C_{\mu }=\partial _{\mu }\theta ^{E}+\frac{(g_{2})^{2}}{g_{3}}(B_{\mu
}^{1}\theta ^{2}-B_{\mu }^{2}\theta ^{1}),  \tag{2.48}  \label{eq:old_2_47}
\end{equation}%
where the parameter convention is $\Theta=g_2\theta^aT_a+g_3\theta^EE$ in
this sector. The invariance condition for $h_{AB}$ guarantees the gauge
invariance of the quadratic Lagrangian.

Spontaneous symmetry breaking in the extended electroweak sector produces
mixing among the neutral fields $W_{\mu }^{3}$, $B_{\mu }$ and $C_{\mu }$.
The detailed mass matrix and its diagonalization were given in~\cite{I1};
only the structural result needed below is recalled here. The
electric-charge operator is not modified by the spontaneous symmetry
breaking 
\begin{equation}
\mathbf{Q}=\mathbf{J}_{3}+\frac{1}{2}\mathbf{T}_{3}.  \tag{2.49}
\label{eq:old_2_48}
\end{equation}%
The central generator $\mathbf{E}$ does not occur in $\mathbf{Q}$. The two
remaining neutral generators can be expressed as 
\begin{equation}
\mathbf{Q}_{Z}=a_{1}\mathbf{J}_{3}+b_{1}\mathbf{T}_{3}+c_{1}\mathbf{E}%
,\qquad \mathbf{Q}_{E}=a_{2}\mathbf{J}_{3}+b_{2}\mathbf{T}_{3}+c_{2}\mathbf{E%
},  \tag{2.50}  \label{eq:old_2_49}
\end{equation}%
with coefficients fixed by the neutral-sector mixing.

Schematically, the symmetry breaking leaves $U(1)_{\mathrm{em}}$ unbroken
and gives masses to the orthogonal gauge directions. Unlike the SM, where $%
U(1)_{Y}$ is introduced as an independent Abelian factor, the
hypercharge-like direction of the present construction is embedded in $%
E^{c}(2)$. The theory therefore also contains the $T_{1},T_{2}$ and central
directions. Their physical spectrum and anomaly constraints must be
specified independently; the existence of the SM charge operator alone does
not establish consistency of the complete extended gauge sector. The gauge
fixing, ghost Lagrangian and neutral-field diagonalization will not be
repeated here and are referred to~\cite{I1}.

2.6 Hypothetical dark-sector candidates

The extended electroweak sector contains possible dark-sector candidates. We
consider states with vanishing weak isospin and hypercharge but nonzero
central charge, $I=0$, $m=0$, and $\kappa\ne0$. These assignments specify
hypothetical internally charged neutral states; they do not by themselves
establish stability, an acceptable abundance, or sufficiently weak
interactions with observable matter. Neutral mixing can induce additional
couplings, which must be assessed phenomenologically.

In the spectral convention of Section 2.3, the corresponding candidate
wavefunctions take the form 
\begin{equation}
h_{n0\kappa}(\xi,\beta)=\sqrt{\frac{|\kappa|}{2\pi}} \exp(-|\kappa|\xi^2/4)%
\exp(i\kappa\beta)L_n^0(|\kappa|\xi^2/2).  \tag{2.51}  \label{eq:old_2_50}
\end{equation}
Here the angular factor is included in the planar normalization, while the
central phase is normalized separately. The proposed identification of these
neutral states as dark matter, and of $n=1,2,3$ as three dark families, is a
hypothesis to be investigated in future work. Their interactions are
discussed in Ref.~\cite{I1}; the associated anomaly conditions are addressed
in Section~4.5.

\bigskip

\textbf{3. Spacetime emergent color SU(3) gauge fields on the TB}

3.1 Threefold multiplicity of colored quark states

The possibility that gauge fields may arise as emergent quantum degrees of
freedom has been investigated in several areas of physics. In the quantum
Hall effect, the low-energy response of a two-dimensional electron system in
an external magnetic field can be described by an effective Chern--Simons
gauge theory, which accounts for the quantized Hall conductivity and
topological edge states \cite{H2,H3}. Emergent and synthetic gauge fields
have subsequently been studied in spin liquids, topological materials,
cold-atom systems, photonic structures, and other strongly correlated
systems. A different mechanism arises from geometric phases: the Berry
connection acts as an effective Abelian gauge potential in parameter space 
\cite{Bery}. Wilczek and Zee showed that, for a degenerate set of quantum
states, this construction generalizes to a non-Abelian gauge connection\cite%
{Wil}. These examples demonstrate that gauge potentials can represent
collective dynamics or the geometry of a quantum-state space rather than
fundamental microscopic gauge fields.

The color group $SU(3)_{C}$ of Quantum Chromodynamics cannot be described as
a geometric symmetry analog as described in section 2. However, these
developments of emergent gauge symmetries and analogies with a Quantum Hall
system motivate the hypothesis that gauge interactions for strong
interaction may be incorporated into the tangent-bundle formulation
following similar ideas.

The construction for emergent $SU(3)_{C}$ gauge fields in the present paper
is the following: The Lie algebra $E^{c}(2)$ determines the vertical
spectral space of the matter fields. The weak $SU(2)$ symmetry arises from a
twofold multiplicity space of chiral fermion states and in a similar way the
color $SU(3)$ symmetry arises from a threefold multiplicity space of quark
states. Thus color is emergent in the following precise sense: $SU(3)$ color
is the compact traceless unitary symmetry acting on three equivalent copies
of the same non-color quark state. The present construction should be
distinguished from the Wilczek--Zee mechanism. In the Wilczek--Zee case, a
non-Abelian connection arises from the adiabatic transport of a degenerate
subspace whose projector varies over a parameter space. This requires both
degeneracy and slowly varying external or background parameters. In the
present construction, by contrast, the starting point is a
representation-theoretic multiplicity. We assume that the same non-color
vertical representation occurs three times in the quark Hilbert space,
giving a three-dimensional copy space. The compact unitary transformations
of this copy space form $U(3)_{C}$, whose traceless part is $SU(3)_{C}$.
Gauging this internal copy symmetry gives the usual color connection.

This mechanism is close in spirit to the formulation of QCD as a Yang--Mills
theory of colored quarks, but the difference in the present tangent-bundle
formulation is that the three color copies are interpreted as a multiplicity
space in the decomposition of the quark Hilbert space, while the vertical
spectral properties are carried by the $E^{c}(2)$ representation space.

The vertical spectral space is spanned by the eigenfunctions $h_{nm\kappa }(%
\mathbf{\xi })$ of the centrally extended Euclidean algebra $e^{c}(2)$ given
in equ.(\ref{eq:old_2_31}) and (\ref{eq:old_2_33}). Here $n,m,\kappa $ are
vertical (internal) quantum numbers. The vertical Hilbert space is%
\begin{equation}
H_{spec}=span\{h_{nm\kappa }(\mathbf{\xi })\}  \tag{3.1}  \label{eq:old_3_1}
\end{equation}

Color generators act on three states with the same electroweak and vertical
quantum numbers. For example, the left-handed quark doublet occurs in three
color copies: $Q_{L}^{r}=(u_{L}^{r},d_{L}^{r})^{T} $, $%
Q_{L}^{g}=(u_{L}^{g},d_{L}^{g})^{T} ,$ $Q_{L}^{b}=(u_{L}^{b},d_{L}^{b})^{T}
. $ Each color copy is itself a weak doublet. Therefore the correct
structure is C$_{\func{col}}^{3}$$\otimes$ C$_{weak}^{2}$.

\bigskip In the multiplicity interpretation, a quark field carries a color
copy index $c=1,2,3.$Thus, schematically,%
\begin{equation}
q_{c}(x,\xi )=\sum_{nm\kappa }q_{c;nm\kappa }(x)h_{nm\kappa }(\xi ). 
\tag{3.2}  \label{eq:old_3_2}
\end{equation}%
where $c=1,2,3$ is the color index. The coefficient fields $q_{cnm\kappa
}(x) $ are ordinary spacetime fields. The color group acts on the index $c$,
not on the fiber variables $\xi$. The generators of $E^{c} (2)$ act on the
vertical functions $h_{nm\kappa }(\mathbf{\xi })$. They do not act on weak
SU(2) or color copy labels. The additional threefold multiplicity $c$ is the
color degree of freedom in the color copy space $\mathbb{C}_{c}^{3}$.

For a left-handed quark doublet the full Hilbert space is%
\begin{equation}
H_{Q;L}=\mathbb{C}_{c}^{3}\otimes \mathbb{C}_{L}^{2}\otimes H_{spec} 
\tag{3.3}  \label{eq:old_3_3}
\end{equation}

For right-handed quarks there is no weak doublet factor, but the color
factor remains. For leptons there is no color factor. The full unitary group
acting on the color copy space $C_{C}^{3}$ is $U(3$)$_{C}$. Its traceless
compact part is $SU(3)_{C}$.Thus $U(3)_C$ is locally described by a common
phase and an $SU(3)_C$ rotation, with a finite identification between their
center elements. The physical nature of the additional Abelian $U(1)_{C}$
factor will be interpreted in the next part of this chapter.

To make the theory invariant under local $SU(3)_{C}$ rotations, one allows
local transformations $U(x)_{C}\in SU(3)_{C}$ acting on the color copy index
by $q(x,\xi )\rightarrow U(x)_{C}q(x,\xi )$ . This is realized by the gauge
principle with the introduction of the spacetime gauge connection%
\begin{equation}
G_{\mu }(x)=G_{\mu }^{A}(x)(\lambda _{A}/2)  \tag{3.4}  \label{eq:old_3_4}
\end{equation}

where $A=1,\ldots,8$, $t^A=\lambda_A/2$, and $\lambda_A$ are the Gell-Mann
matrices, normalized by $\func{tr}(t^At^B)=\delta^{AB}/2$. The
color-covariant derivative is%
\begin{equation}
D_{\mu }=\partial _{\mu }-ig_{s}G_{\mu }^{A}(x)(\lambda _{A}/2)  \tag{3.5}
\label{eq:old_3_5}
\end{equation}

The SU(3) field strength is then the curvature of this spacetime connection:

\begin{equation}
G_{\mu \nu }=\partial _{\mu }G_{\nu }-\partial _{\nu }G_{\mu }-ig_{s}[G_{\mu
},G_{\nu }]  \tag{3.6}  \label{eq:old_3_6}
\end{equation}%
.

In components this is%
\begin{equation}
G_{\mu \nu }^{A}=\partial _{\mu }G_{\nu }^{A}-\partial _{\nu }G_{\mu
}^{A}+g_{s}f^{ABC}G_{\mu }^{B}G_{\nu }^{C}  \tag{3.7}  \label{eq:old_3_7}
\end{equation}

Here $f^{ABC}$ are the structure constants of $SU(3)$.

The gluon part of the Lagrangian is%
\begin{equation}
L_{gluon}=-\frac{1}{4}G_{\mu \nu }^{A}G^{A\mu \nu }  \tag{3.8}
\label{eq:old_3_8}
\end{equation}

Quarks are described by a projected spinor field with an assumed local Dirac
kinetic term. The projection alone does not derive the Dirac equation. Let $%
S $ denote the spin bundle over spacetime. The matter field is a section 
\begin{equation}
q(x,\mathbf{\xi })\in S_{x}\otimes \mathcal{H}_{\mathrm{low}}.  \tag{3.9}
\label{eq:old_3_9}
\end{equation}%
In components, 
\begin{equation}
q(x,\mathbf{\xi })=\sum_{c=1}^{3}q_{c}(x)\otimes |\chi _{c}(\mathbf{\xi }%
)\rangle ,  \tag{3.10}  \label{eq:old_3_10}
\end{equation}%
where $|\chi _{c}(\mathbf{\xi })\rangle $ is an internal basis state and
each $q_{c}(x)$ is a Dirac spinor. We use $q(x)=(q_{1},q_{2},q_{3})^{T}$ for
the projected four-dimensional quark field. Its kinetic term is%
\begin{equation}
L_{quark}=\bar{q}(i\gamma ^{\mu }D_{\mu }-M_{q})q  \tag{3.11}
\label{eq:old_3_11}
\end{equation}

In the tangent-bundle approach the quark mass matrix $M_{q}$ should not be
regarded as fundamental. Here it is an effective parameter of the broken
low-energy theory, not a bare mass term in the unbroken chiral electroweak
theory. It is expected to arise dynamically from internal vacuum structures,
especially from interactions with the $n=0$ sector.

The vertical $E^{c}(2)$ sector is described by internal variables and by a
Landau-level-like spectrum. The internal excitations in the ground state $%
n=0 $ are assumed to carry an internal vertical charge density. This does
not assert a nonzero spacetime charge density. At first sight this may
appear to introduce a preferred direction with a violation of Lorentz
invariance, because the representation of this group resembles a
nonrelativistic planar spectral problem. However, this conclusion would be
incorrect.

The variables of the $E^{c}(2)$ sector are vertical internal variables, not
physical spacetime coordinates. Therefore the internal Landau-level
structure does not by itself select a preferred spacetime frame. The color
construction is Lorentz compatible because its multiplicity space is
internal. The Lorentz properties of a dynamical vacuum condensate require a
separate analysis. Lorentz transformations act on spacetime and spinor
indices, whereas $SU(3)$ color acts on the color copy index. These actions
commute.

The gluon potential $G_{\mu}^{A}$ is a spacetime one-form and transforms as
a Lorentz vector in its spacetime index $\mu$, while the adjoint index $A$
is internal. The field strength $G_{\mu\nu}^{A}$ is a Lorentz tensor in $\mu$
and $\nu$, and the contraction $G_{\mu\nu}^{A}G^{A\mu\nu}$ is a Lorentz
scalar.

The Dirac Lagrangian is Lorentz invariant because $\gamma^{\mu}D_{\mu}$ is
the covariant Dirac operator. In a tetrad formulation one writes $%
\gamma^{\mu}(x)=e_{a}^{\mu}(x)\gamma^{a}$.

Hence the QCD Lagrangian is Lorentz invariant: $SU(3)_{\mathrm{col}}$ acts
on an internal copy index and commutes with Lorentz transformations.

\bigskip 3.2 Global structure and the color center

Section~3.1 identifies the local color generators. Their global group also
depends on which finite transformations are regarded as identical \cite%
{Tong,Aha,H.Lin}. An electromagnetic transformation acts as a scalar phase
on a fixed-charge color triplet. This alone does not determine a common
kernel, meaning a set of combined color and phase transformations that act
as the identity on the represented states. We specify that identification by
the construction below.

First consider rotations $R(\theta )$ generated by the reference operator $%
T_{3}^{(0)}$ in (\ref{eq:old_2_29}). On the ordinary plane $R(2\pi )=1$. In
the internal quantum group we allow three turns before identifying a
rotation with the identity: 
\begin{equation}
R(\theta )=\exp (i\theta T_{3}),\qquad R(6\pi )=1,\qquad R(2\pi )^{3}=1. 
\tag{3.12}  \label{eq:rotation_period}
\end{equation}%
This is a threefold cover of the rotation subgroup: angles differing by $%
2\pi $ have the same action on planar vectors, but can give different phases
on quantum states. The generator in the reference sector is $%
T_{3}=T_{3}^{(0)}$; its phase-shifted representations are given below. We
denote the resulting centrally extended internal group by $E^{c}(2)_{\mathrm{%
rot}(6\pi )}$. The suffix specifies the rotation period, not a new spatial
dimension or a third translation generator. The independent central
coordinate $\beta $ still has period $2\pi $.

The element $R(2\pi)$ commutes with every group transformation and is
therefore central. In the covering group its first and second powers need
not be the identity, while its third power is the identity by (\ref%
{eq:rotation_period}); this is the meaning of order three. The color center
consists of 
\begin{equation}
Z(SU(3)_c)=\{\mathbf{1}_3,\omega\mathbf{1}_3,\omega^2\mathbf{1}_3\}, \qquad
\omega=e^{2\pi i/3}.  \tag{3.13}  \label{eq:color_center}
\end{equation}
We identify a simultaneous color-center rotation and an opposite vertical $%
2\pi$ rotation with the identity: 
\begin{equation}
G_{\mathrm{vert}}= \frac{SU(3)_c\times SU(2)_L\times E^c(2)_{\mathrm{rot}%
(6\pi)}} {\langle(\omega\mathbf{1}_3,\mathbf{1}_2,R(-2\pi))\rangle}. 
\tag{3.14}  \label{eq:global_group}
\end{equation}
The angle brackets mean the three transformations obtained by taking the
zeroth, first and second powers of the displayed element. Dividing by this
set means that pairs differing by one of these transformations are
identified. No local generator or new gauge field is added by this step. The
rotation cover is a choice for the internal quantum group, not a threefold
cover of the full Lorentz frame group.

3.3 Fractional charges and the rotation phases

A phase-shifted rotation representation is obtained from the reference one
by 
\begin{equation}
U_{\sigma }(R(\theta ))=e^{i\sigma \theta }U_{0}(R(\theta )),\qquad \mathbf{T%
}_{3}=\mathbf{T}_{3}^{(0)}+\sigma \mathbf{1}.  \tag{3.15}
\label{eq:shift_generator}
\end{equation}%
The condition $R(6\pi )=1$ requires $3\sigma \in \mathbb{Z}$. Since $\mathbf{%
T}_{3}^{(0)}h_{nm\varkappa }=mh_{nm\varkappa _{{}}}$ for the functions in (%
\ref{eq:old_2_31}), 
\begin{equation}
\mathbf{T}_{3}h_{nm\varkappa _{{}}}=Yh_{nm\varkappa _{{}}},\qquad Y=m+\sigma
,\qquad m\in \mathbb{Z}.  \tag{3.16}  \label{eq:Yshift}
\end{equation}%
The functions $h_{nm\varkappa _{{}}}$retain their integer orbital factor $%
e^{im\varphi }$. All internal quantum numbers, including the physical
interpretation of $Y$ and $m$, are collected in Section~4.2.

Let $r=0,1,2$ specify the color-center phase $\omega^r$ of a representation.
In particular $r=0$ for a singlet and $r=1$ for a triplet. The
transformation identified in (\ref{eq:global_group}) then acts as $e^{2\pi
ir/3}e^{-2\pi i\sigma}$. Requiring this phase to be one gives 
\begin{equation}
\sigma\equiv r/3\pmod1.  \tag{3.17}  \label{eq:sigma_triality}
\end{equation}
Thus quark triplets may use $\sigma=1/3$, while leptons use $\sigma=0$.

To select the observed representative of the quark class, use the mixed
weak-hypercharge anomaly condition derived in (\ref{eq:SU2T3}), $%
3Y_{Q}+Y_{L}=0$. The lepton normalization $Y_{L}=-1$ follows from the
measured neutrino and electron charges together with $\mathbf{Q}=\mathbf{J}%
_{3}+Y/2$: for the neutrino $0=1/2+Y_{L}/2$, and for the electron $%
-1=-1/2+Y_{L}/2$. For $Y_{Q}=m_{Q}+1/3$ the anomaly condition becomes $%
3m_{Q}=0$, giving $m_{Q}=0$ and $Y_{Q}=1/3$. Equal electric charges of the
left- and right-handed components of each quark require $Y_{u_{R}}=Y_{Q}+1$
and $Y_{d_{R}}=Y_{Q}-1$. The same reasoning gives $Y_{e_{R}}=-2$. With the
selected SM fermion content and no additional contribution to this mixed
anomaly, the assignments are therefore 
\begin{equation}
\begin{array}{c|rrrrr}
& Q_{L} & u_{R} & d_{R} & L_{L} & e_{R} \\ \hline
m & 0 & 1 & -1 & -1 & -2 \\ 
\sigma  & 1/3 & 1/3 & 1/3 & 0 & 0 \\ 
Y & 1/3 & 4/3 & -2/3 & -1 & -2%
\end{array}%
.  \tag{3.18}  \label{eq:observable_assignments}
\end{equation}%
This argument fixes charge assignments, without choosing the Higgs
mass-generation mechanism.

The surviving electromagnetic generator is 
\begin{equation}
\mathbf{Q}=\mathbf{J}_{3}+\frac{1}{2}\mathbf{T}_{3}.  \tag{3.19}
\label{eq:Qphysical}
\end{equation}%
The extra bosons $B^{\pm }$ originate from the connection of $\mathbf{T}_{1},%
\mathbf{T}_{2}$. Their electric charge follows directly from 
\begin{equation}
\lbrack \mathbf{J}_{3},\mathbf{T}_{\pm }]=0,\qquad \lbrack \mathbf{T}_{3},%
\mathbf{T}_{\pm }]=\pm \mathbf{T}_{\pm },\qquad \lbrack \mathbf{Q},\mathbf{T}%
_{\pm }]=\pm \frac{1}{2}\mathbf{T}_{\pm }.  \tag{3.20}  \label{eq:Bcharge}
\end{equation}%
Indeed, $\mathbf{T}_{\pm }$ takes a state of charge $q$ into one of charge $%
q\pm 1/2$, whenever the matrix element is nonzero. The corresponding
conjugate boson must carry the compensating charge at a vertex; the $%
B^{+},B^{-}$ pair has charges of magnitude $1/2$. Which connection
coefficient is called $B^{+}$ depends on the convention for writing the
conjugate pair in the interaction. The additional $C_{\mu }$ potential is
associated with $\mathbf{E}$ and has zero electromagnetic adjoint charge
because $[\mathbf{Q},\mathbf{E}]=0$; after symmetry breaking it enters the
neutral-field mixture.

For completeness, the global electromagnetic phase can now be normalized
explicitly. Let $\alpha $ be the dimensionless parameter in $\exp (i\alpha 
\mathbf{Q})$. Since the factors commute, 
\begin{equation}
\exp (i\alpha \mathbf{Q})=\exp (i\alpha \mathbf{J}_{3})\exp (i\alpha \mathbf{%
T}_{3}/2).  \tag{3.21}  \label{eq:em_rotation_factors}
\end{equation}%
The weak rotation parameter is therefore $\alpha $, and the vertical
rotation parameter is $\alpha /2$. These are group parameters, not the
Weinberg mixing angle of the gauge fields. The first factor has period $4\pi 
$ on a weak doublet, and the second has period $12\pi $ in $\alpha $ by (\ref%
{eq:rotation_period}). Define 
\begin{equation}
\mathbf{N}=6\mathbf{Q},\qquad \varphi _{\mathrm{em}}=\alpha /6,\qquad \exp
(i\alpha \mathbf{Q})=\exp (i\varphi _{\mathrm{em}}\mathbf{N}).  \tag{3.22}
\label{eq:Nnormalization}
\end{equation}%
Now $\varphi _{\mathrm{em}}$ has period $2\pi $ and $\mathbf{N}$ has integer
eigenvalues. This is a normalization of the same electromagnetic group, not
a new charge or particle label. The factor six accommodates both quark
charges in thirds and $B^{\pm }$ charges in halves: for example $N_{u}=4$, $%
N_{d}=-2$, $N_{e}=-6$ and $N_{B^{\pm }}=\pm 3$.

At $\alpha=4\pi$, (\ref{eq:em_rotation_factors}) has weak factor one and
vertical factor $R(2\pi)$. Using (\ref{eq:global_group}), the surviving
group can be written 
\begin{equation}
G_{c\mathrm{em}}= \frac{SU(3)_c\times U(1)_N} {\langle(\omega\mathbf{1}%
_3,e^{-2\pi i/3})\rangle}.  \tag{3.23}  \label{eq:cem_group}
\end{equation}
On a state of triality $r$ and integer charge $N$, the identified element
has phase $\exp[2\pi i(r-N)/3]$. Requiring it to be one gives 
\begin{equation}
N\equiv r\pmod3,\qquad Q=\frac r6+\frac k2,\quad k\in\mathbb{Z}.  \tag{3.24}
\label{eq:charge_class}
\end{equation}
Here $k$ simply enumerates the allowed representatives. The specific
observable fermions are those in (\ref{eq:observable_assignments}).

\bigskip\textbf{4. Quantization on the tangent bundle, BRST structure and
anomaly cancellation}

4.1 Quantization of gauge fields with non-semisimple Lie groups

The standard perturbative quantization of internal gauge fields relies on an
invariant kinetic form, gauge fixing, and Faddeev--Popov ghosts. While
compact semisimple groups combined with Abelian factors provide the standard
setting, semisimplicity alone is not a necessary condition, as $U(1)$
illustrates. For a non-semisimple algebra, the Killing form can be
degenerate; consequently, an alternative nondegenerate invariant form must
be specified. The central extension $e^c(2)$ admits such a form, (\ref%
{eq:old_2_42}), allowing for an action and formal covariant quantization 
\cite{Ban,Tsey}. The existence of this action must be assessed independently
from physical-state unitarity.

Matter-state quantization uses unitary representations of the internal gauge
group on a Hilbert space. Quantum symmetries act on rays and may require
projective representations whose operators satisfy 
\begin{equation}
U(g_{1})U(g_{2})=e^{i\alpha (g_{1},g_{2})}U(g_{1}g_{2}),  \tag{4.1}
\label{eq:projective_composition}
\end{equation}%
rather than an ordinary composition law \cite{Wig,Bar,Wein}. For spatial
rotations these representations are accommodated by passing from $SO(3)$ to
its double cover $SU(2)$, which admits half-integer spin without changing
the local Lie algebra. For the nontrivial projective sector chosen for $E(2)$%
, its universal cover is not enough: one uses a central extension
incorporating the additional $U(1)$ phase. A cover and a Lie-algebra central
extension solve different problems. The former changes global topology; the
latter replaces $[\mathbf{T}_{1},\mathbf{T}_{2}]=0$ by $[\mathbf{T}_{1},%
\mathbf{T}_{2}]=i\mathbf{E}$ in (\ref{eq:old_2_19}).

The invariant internal kinetic form (\ref{eq:old_2_42}) is indefinite. This
concerns the gauge-field action, not a requirement that the Casimir have
positive eigenvalues: the sign of a Casimir spectrum alone does not
determine unitarity of a representation. Quantization with an indefinite
metric in the enlarged state space is not inconsistent by itself, as
covariant quantum electrodynamics demonstrates. For $E^c(2)$, however, the
negative internal kinetic direction also involves transverse vector modes.
Consequently, introducing ghosts does not automatically prove that the BRST
physical subspace has positive norm. Ref.~\cite{Tsey} discusses this issue
and suggests a possible reinterpretation through a decoupled sector
compatible with unitarity; it does not establish the corresponding
matter-coupled construction here. The gauge fixing and ghost treatment of
Ref.~\cite{I1} provide the starting point.

The noncompact group can require continuous labels across representation
sectors and regulated traces.This does not prevent a countable orthonormal
basis in a fixed nonzero-$\kappa $ sector. Normalization, completeness and
projection onto four-dimensional fields must account for the sectors
retained. Other model-dependent constructions are discussed in Refs.~\cite%
{Na,Fer}.

4.2 Quantization of matter fields on the tangent bundle

The local reduced Lie-algebra is $su(2)\oplus e^{c}(2)$. Matter fields are
sections of an associated Hilbert bundle and are expanded in the vertical
basis $h_{f}(\mathbf{\xi })$ of (\ref{eq:old_2_33}). We collect the internal
labels as 
\begin{equation}
f=(n,m,I,I_{3},\kappa ,\sigma ).  \tag{4.2}  \label{eq:internal_labels}
\end{equation}%
Their physical meanings are as follows. The nonnegative integer $n$ is the
radial spectral index. The model associates $n=0$ with the proposed internal
vacuum and assigns higher levels to particle families; selecting three
observable families requires dynamics. The integer $m$ is the eigenvalue of
the reference angular operator $\mathbf{T}_{3}^{(0)}$, not necessarily the
hypercharge. The labels $I$ and $I_{3}$ are weak isospin and its third
component: observable left-handed fermions have $I=1/2$ and $I_{3}=\pm 1/2$,
while right-handed singlets have $I=I_{3}=0$. The central charge $\kappa $
is the eigenvalue of $E=-i\partial _{\beta }$; it is integer for central
period $2\pi $. The offset $\sigma $ specifies the global rotation
representation in (\ref{eq:shift_generator}); it is fixed for a
representation and is not a further excitation index.

The physical weak hypercharge is the eigenvalue of $T_3$: 
\begin{equation}
Y=m+\sigma,\qquad Q=I_3+Y/2.  \tag{4.3}  \label{eq:physical_labels}
\end{equation}
For quark triplets $\sigma=1/3$ can be chosen, so $m=0,1,-1$ yields $%
Y=1/3,4/3,-2/3$ for the assignments in (\ref{eq:observable_assignments}).
For leptons $\sigma=0$ and $Y=m$. Color is a separate three-component copy
index $c$, suppressed in $f$. The continuous variables $\xi_1,\xi_2$
describe positions in the internal plane, not additional spacetime
coordinates or discrete particle charges.

The shifted Casimir has the same radial functions as the reference one, but
eigenvalues 
\begin{equation}
\mathbf{\Delta }_{\sigma }=\mathbf{\Delta }_{0}+2\sigma \mathbf{E},\qquad
\epsilon _{nm\kappa ,\sigma }=|\kappa |(2n+|m|+1)+\kappa m+2\sigma \kappa . 
\tag{4.4}  \label{eq:shifted_spectrum}
\end{equation}%
The centrifugal term still contains $m^{2}/\xi ^{2}$ with $m=Y-\sigma $. The
global shift does not by itself fix $\kappa $ in terms of $Y$; the
conditional anomaly selection is derived in Section~4.5.

The fermion field depends on both spacetime coordinates and vertical
coordinates: $\Psi =\Psi (x,\mathbf{\xi }).$ In the Minkowski spacetime base
manifold it can be expanded as 
\begin{equation}
\Psi (x,\mathbf{\xi })=\sum_{f,s}\int \frac{d^{3}p}{(2\pi )^{3}}\frac{1}{%
\sqrt{2\epsilon _{p}}}\left[ e^{-ip\cdot x}\mathbf{a}_{fs}(p)u_{s}(p)h_{f}(%
\mathbf{\xi })+e^{ip\cdot x}\mathbf{b}_{fs}^{\dagger }(p)v_{s}(p)h_{f}^{\ast
}(\mathbf{\xi })\right] .  \tag{4.5}  \label{eq:old_4_1}
\end{equation}

Here $u_{s}(p)$\ with $p=(\epsilon (\mathbf{p}),\mathbf{p})$ is the
plane-wave solution of the Dirac equation for a particle and $v_{s}(p)$\
that of an anti-particle, $p\cdot x$ denotes the spacetime scalar product.
We use continuum normalization for momenta, $s$\emph{\ }is the helicity\emph{%
.}The adjoint operator $\overline{\Psi (x)}$ is defined in the standard way: 
$\overline{\Psi (x)}=\Psi ^{\dagger }(x)\gamma ^{0}.$The creation and
annihilation operators satisfy 
\begin{equation}
\{\mathbf{a}_{fs}(p),\mathbf{a}_{f^{\prime }s^{\prime }}^{\dagger
}(p^{\prime })\}=(2\pi )^{3}\delta ^{(3)}(p-p^{\prime })\delta _{ff^{\prime
}}\delta _{ss^{\prime }},  \tag{4.6}  \label{eq:old_4_2}
\end{equation}

and 
\begin{equation}
\{\mathbf{b}_{fs}(p),\mathbf{b}_{f^{\prime }s^{\prime }}^{\dagger
}(p^{\prime })\}=(2\pi )^{3}\delta ^{(3)}(p-p^{\prime })\delta _{ff^{\prime
}}\delta _{ss^{\prime }}.  \tag{4.7}  \label{eq:old_4_3}
\end{equation}

All other anticommutators vanish. The orthonormality relation for $h_{f}(%
\mathbf{\xi })$ is 
\begin{equation}
\int d\mu (\mathbf{\xi })\,h_{f}^{\ast }(\mathbf{\xi })h_{f^{\prime }}(%
\mathbf{\xi })=\delta _{ff^{\prime }}.  \tag{4.8}  \label{eq:old_4_4}
\end{equation}%
For a complete planar basis in a fixed internal representation sector, and
the corresponding planar measure, the completeness relation is 
\begin{equation}
\sum_{f}h_{f}(\mathbf{\xi })h_{f}^{\ast }(\mathbf{\xi }^{\prime })=\delta
(\xi _{1}-\xi _{1}^{\prime })\delta (\xi _{2}-\xi _{2}^{\prime }).  \tag{4.9}
\label{eq:old_4_5}
\end{equation}%
A truncation to selected modes replaces the full basis sum by the projection
onto those modes. Using the complete kernel and standard Dirac spinor
normalization, one obtains 
\begin{equation}
\{\Psi _{\alpha }(t,\mathbf{x},\mathbf{\xi }),\Psi _{\beta }^{\dagger }(t,%
\mathbf{y},\mathbf{\xi }^{\prime })\}=\delta _{\alpha \beta }\delta ^{(3)}(%
\mathbf{x}-\mathbf{y})\delta (\mathbf{\xi }-\mathbf{\xi }^{\prime }). 
\tag{4.10}  \label{eq:old_4_6}
\end{equation}%
Projection onto one vertical mode gives 
\begin{equation}
\Psi _{f}(x)=\int d\mu (\mathbf{\xi })\,h_{f}^{\ast }(\mathbf{\xi })\Psi (x,%
\mathbf{\xi }).  \tag{4.11}  \label{eq:old_4_7}
\end{equation}

Thus the projected field satisfies the standard rules 
\begin{equation}
\{\Psi _{f\alpha }(t,\mathbf{x}),\Psi _{f^{\prime }\beta }^{\dagger }(t,%
\mathbf{y})\}=\delta _{\alpha \beta }\delta _{ff^{\prime }}\delta ^{(3)}(%
\mathbf{x}-\mathbf{y}).  \tag{4.12}  \label{eq:old_4_8}
\end{equation}%
\begin{equation}
\{\Psi _{f}(t,\mathbf{x}),\Psi _{f{\acute{}}}(t,\mathbf{y})\}=\{\Psi
_{f}^{\dagger }(t,\mathbf{x}),\Psi _{f{\acute{}}}^{\dagger }(t,\mathbf{y}%
)\}=0  \tag{4.13}  \label{eq:old_4_9}
\end{equation}%
These relations are understood at equal times. The conjugate-mode convention
in the antiparticle term of (\ref{eq:old_4_1}) must be matched to the
particle and antiparticle internal bases before applying the projection.

Matter quantization in equ.(\ref{eq:old_4_1}) uses the vertical
eigenfunctions $h_{f}(\mathbf{\xi })$ explicitly. The labels $f$ are not
arbitrary flavor labels; they are determined by the vertical spectral
problem as eigenfunctions of the Casimir operator.

The projected four-dimensional Green functions have the standard form, while
their internal labels are those in (\ref{eq:internal_labels}). The unpaired
vacuum $\mid 0\succ $ is defined such that 
\begin{equation}
\mathbf{a}_{f,s}(\mathbf{p})\mid 0\succ =0,\qquad \mathbf{b}_{f,s}(\mathbf{p}%
)\mid 0\succ =0  \tag{4.14}  \label{eq:old_4_10}
\end{equation}

The creation operator adds one particle with quantum number $p$ in the
state. Thus the single one-particle states are spanned by 
\begin{equation}
|\mathbf{p},s,f\succ =\mathbf{a}_{f,s}^{\dagger }(\mathbf{p})|0\succ 
\tag{4.15}  \label{eq:old_4_11}
\end{equation}

This generalizes the standard presentation of Quantum Field Theory (see e.g. 
\cite{Wein}) in which one actually starts from relativistic particle states
and only later introduces quantum fields. In this representation single
particle states are defined as states in a Hilbert space carrying one
unitary irreducible projective representation of the Poincar\'{e} group.
Quantum fields in the SM carry the internal Quantum number isospin and
hypercharge in electroweak interaction and color in strong interaction
(QCD). Quantum fields on the tangent bundle carry in addition the discrete
eigennumbers of the vertical Laplacian the family IQN $n$, the $E^{c}$
charge $\varkappa $ and the continuous variable $\mathbf{\xi }.$ Note that
in this description of the states we don't have the spinors $u(\mathbf{p})$
and $v(\mathbf{p})$. The spin is captured by the label $s=\pm 1/2$ of the
basis states $|\mathbf{p},s,f\succ $ =$\mathbf{a}_{f,s}^{\dagger }\mid
0\succ .$ Based on such single particle Hilbert space one can then carry out
the Fock space construction and build a Hilbert space whose states are all
possible multiparticle states. In such a way the field is constructed out of 
$\mathbf{a}_{M,s}^{\dagger }(\mathbf{p})$ and $\mathbf{a}_{M,s}(\mathbf{p}).$

The construction therefore separates three steps: solving the internal
spectral problem, quantizing the spacetime coefficient fields, and
restricting their charges by gauge consistency. The internal quantum numbers
have the meanings given in (\ref{eq:internal_labels}) and (\ref%
{eq:physical_labels}).

4.3 Quantization of gauge fields on the tangent bundle

As described in section3 QCD gauge fields and Dirac matter arises in the TB
as emergent structures with semi-simple $SU(3)$ symmetry in (3+1)D
spacetime. Therefore the quantization of these fields can be formulated as
in the SM.

Gauge potentials and gauge-field Lagrangian for the extended electroweak
sector are discussed in section 2.4. The gauge fields are
spacetime-dependent connection coefficients multiplying internal generators:$%
\mathcal{W}_{\mu }(x)=W_{\mu }^{A}(x)\mathbf{T}_{A},$ and are not expanded
like matter fields in the eigenfunctions $h_{nm\kappa }(\mathbf{\xi })$. The
internal quantum numbers of matter fields arise from the representation
space on which the generators $\mathbf{T}_{A}$ act. The gauge fields
themselves are geometric connections on this internal bundle. In the reduced
electroweak sector the connection is 
\begin{equation}
\mathcal{W}_{\mu }=g_{1}W_{\mu }^{i}\mathbf{J}_{i}+g_{2}B_{\mu }^{a}\mathbf{T%
}_{a}+g_{3}C_{\mu }\mathbf{E}.  \tag{4.16}  \label{eq:old_4_12}
\end{equation}%
This is the connection of (\ref{eq:old_2_37}), before neutral mixing. After
symmetry breaking the photon couples to $\mathbf{Q}=\mathbf{J}_{3}+\mathbf{T}%
_{3}/2$, not to $\mathbf{T}_{3}$ alone.

The gauge-field path integral is 
\begin{equation}
Z=\int \mathcal{D}\mathcal{W}\,\mathcal{D}\bar{\Psi}\,\mathcal{D}\Psi \,\exp
(iS).  \tag{4.17}  \label{eq:old_4_13}
\end{equation}

Because the connection has gauge redundancy, gauge fixing is required.
Choose gauge-fixing functions 
\begin{equation}
\mathcal{G}^{A}[\mathcal{W}]=\partial^\mu W_{\mu }^{A}.  \tag{4.18}
\label{4.14}
\end{equation}

Here $W$ $_{\mu }^{A}$ denotes the component fields $A_{\mu },\quad W_{\mu
}^{1},\quad W_{\mu }^{2},\quad C_{\mu }$ .

The gauge-fixed path integral contains the Faddeev--Popov determinant

\begin{equation}
\det \left( \frac{\delta \mathcal{G}^{A}[\mathcal{W}^{\alpha }]}{\delta
\alpha ^{B}}\right) .  \tag{4.19}  \label{eq:old_4_15}
\end{equation}

This determinant is represented by ghost and antighost fields $c^{A},\bar{c}%
_{A}.$

The ghost fields are algebra-valued spacetime fields: 
\begin{equation}
c(x)=c^{A}(x)\mathbf{T}_{A}.  \tag{4.20}  \label{eq:old_4_16}
\end{equation}

They are not expanded in vertical eigenfunctions $h_{f}(\mathbf{\xi })$.
Like the gauge connection, they belong to the gauge redundancy of the
internal connection, not to the matter Hilbert basis. The Lagrangian of
gauge fields with gauge fixing term and Faddeev-Popov ghosts is given by 
\begin{equation}
\mathcal{L}_{\mathrm{gauge}}=-\frac{1}{4}g_{AB}F_{\mu \nu }^{A}F^{B\mu \nu }-%
\frac{1}{2\alpha }\partial ^{\mu }A_{\mu }^{A}\partial ^{\nu }A_{\nu }^{A}+%
\mathcal{L}_{\mathrm{gh}}  \tag{4.21}  \label{eq:old_4_17}
\end{equation}

with the invariant form $g_{AB}$ and coupling normalization specified in
Section 2.5. An auxiliary-field formulation fixes the gauge-fixing and ghost
sign conventions as in Section 4.4. The ghost Lagrangian $\mathcal{L}_{%
\mathrm{gh}}$ has the schematic form 
\begin{equation}
\mathcal{L}_{\mathrm{gh}}=-\bar{c}_{A}\partial ^{\mu }(D_{\mu }c)^{A}. 
\tag{4.22}  \label{eq:old_4_18}
\end{equation}

\bigskip 4.4 BRST invariance in the reduced TB quantum field theory

In the standard quantization of non-Abelian gauge fields (in particular QCD)
the effective Lagrangian has an extended symmetry with transformations
denoted as BRST transformations \cite{BRST} which plays a crucial role in
proving the consistency of the quantized framework of the gauge theory.The
BRST invariance is essential for the proofs of unitarity, gauge independence
(after gauge fixing) and renormalizability of gauge theories. Let us
consider whether the same invariance can be proven for the derived emergent
QCD and in the reduced tangent bundle formalism.

For the projected color sector, use $D_\mu q=(\partial_\mu-ig_sG_\mu^At^A)q$%
. With an auxiliary Nakanishi--Lautrup field $B^A$, a consistent gauge-fixed
Lagrangian is 
\begin{align}
\mathcal{L}_{SU(3)}={}&-\frac14G_{\mu\nu}^AG^{A\mu\nu} +\bar q(i\gamma^\mu
D_\mu-M_q)q  \notag \\
&+B^A\partial^\mu G_\mu^A+\frac\alpha2B^AB^A -\bar c^A\partial^\mu(D_\mu
c)^A.  \tag{4.23}  \label{eq:old_4_19}
\end{align}
Here $M_q$ is an effective color-singlet mass matrix in the broken theory,
and 
\begin{equation}
(D_\mu c)^A=\partial_\mu c^A+g_sf^A{}_{BC}G_\mu^Bc^C.  \tag{4.24}
\label{eq:old_4_20}
\end{equation}
Eliminating $B^A$ gives 
\begin{equation}
\mathcal{L}_{\mathrm{gf}}=-\frac1{2\alpha} (\partial^\mu
G_\mu^A)(\partial^\nu G_\nu^A).  \tag{4.25}  \label{eq:old_4_21}
\end{equation}
Define the Grassmann-odd BRST operator $s$ by 
\begin{equation}
sG_\mu^A=(D_\mu c)^A.  \tag{4.26}  \label{eq:old_4_22}
\end{equation}
\begin{equation}
sc^A=-\frac{g_s}{2}f^A{}_{BC}c^Bc^C.  \tag{4.27}  \label{eq:old_4_23}
\end{equation}
\begin{align}
s\bar c^A&=B^A,  \tag{4.28}  \label{eq:old_4_24} \\
sB^A&=0.  \tag{4.29}  \label{eq:old_4_25}
\end{align}
For matter, $sq=ig_sc^At^Aq$ and $s\bar q=ig_s\bar q c^At^A$, with Grassmann
order fixed as written. The Jacobi identity yields $s^2=0$. The gauge-fixing
and ghost terms are $s[\bar c^A(\partial^\mu G_\mu^A+\alpha B^A/2)]$.

Color labels equivalent copies rather than three different scalar vertical
wavefunctions. For a fixed vertical representation, use 
\begin{equation}
q(x,\mathbf{\xi })=\sum_{c=1}^{3}q_{c}(x)|c\rangle \otimes h_{\lambda }(%
\mathbf{\xi }),\qquad \langle c|d\rangle =\delta _{cd}.  \tag{4.30}
\label{eq:old_4_26}
\end{equation}%
The color generators act on $|c\rangle $ and leave $h_{\lambda }$ unchanged.
Thus 
\begin{equation}
sq(x,\mathbf{\xi })=\sum_{c}(ig_{s}c^{A}t_{cd}^{A}q_{d}(x))|c\rangle \otimes
h_{\lambda }(\mathbf{\xi }).  \tag{4.31}  \label{eq:old_4_27}
\end{equation}%
Projection using the copy-space inner product and $\int d\mu (\mathbf{\xi }%
)|h_{\lambda }(\mathbf{\xi })|^{2}=1$ gives 
\begin{equation}
sq_{c}(x)=ig_{s}c^{A}t_{cd}^{A}q_{d}(x).  \tag{4.32}  \label{eq:old_4_28}
\end{equation}%
The projected color BRST complex is therefore the ordinary semisimple $SU(3)$
complex. Its Slavnov--Taylor identities constrain the counterterms of the
local QCD action. This result does not prove unitarity or anomaly freedom of
the complete non-semisimple electroweak extension.

4.5 Anomaly Cancellation

In the SM anomalies arise because a symmetry that holds at the classical
level can fail after quantization, especially for chiral gauge currents in
fermion triangle loop diagrams. In the Standard Model, the dangerous gauge
anomalies are canceled by a precise balance of quark and lepton charges and
representations within each generation, so the theory remains
gauge-consistent.The cancellation is not accidental: the hypercharges of
left- and right-handed fermions are arranged so that the contributions from
all fermions sum to zero. In practice, the quark sector and lepton sector
cancel each other's gauge-anomaly contributions, and this works generation
by generation in the Standard Model \cite{Wein,Pes}.

Defining the currents of the generator $X_{A}$ at spacetime points $x$:%
\begin{equation}
j_{A}^{\mu }(x)=\int d\mathbf{\xi }\overline{\psi }(x,\mathbf{\xi })\gamma
^{\mu }X_{A}\psi (x,\mathbf{\xi })  \tag{4.33}  \label{eq:old_4_29}
\end{equation}%
and the corresponding axial-vector current%
\begin{equation}
j_{A5}^{\mu }(x)=\int d\mathbf{\xi }\overline{\psi }(x,\mathbf{\xi })\gamma
^{\mu }\gamma _{5}X_{A}\psi (x,\mathbf{\xi })  \tag{4.34}
\label{eq:old_4_30}
\end{equation}%
Using Wick's theorem to contract the fermion fields, the triangle diagram
splits into two distinct orientations (the direct loop and the crossed loop
to ensure Bose symmetry under y$\longleftrightarrow $ z, $\alpha
\longleftrightarrow \beta )$. Computing the chiral anomaly triangle diagram
involves evaluating the three-point correlation function of one axial-vector
current and two vector currents. The three-point correlation function is 
\begin{equation}
\Gamma _{ABC}^{\mu \alpha \beta }=\prec 0\mid T\{j_{A5}^{\mu
}(x)j_{B}^{\alpha }(y)j_{C}^{\beta }(z)\}\mid 0\succ  \tag{4.35}
\label{eq:old_4_31}
\end{equation}

For a general chiral gauge theory with generators $\mathbf{X}_{A}$, the
local triangle anomaly is governed by the symmetrized chiral trace%
\begin{equation}
\mathcal{A}_{ABC}=\sum_{\mathrm{left}}\func{tr}_{R}\!\left( \{\mathbf{X}_{A},%
\mathbf{X}_{B}\}\mathbf{X}_{C}\right) -\sum_{\mathrm{right}}\func{tr}%
_{R}\!\left( \{\mathbf{X}_{A},\mathbf{X}_{B}\}\mathbf{X}_{C}\right) . 
\tag{4.36}  \label{eq:old_4_32}
\end{equation}%
Equivalently, if all fields are written as left-handed Weyl fields, the
right-handed fields are replaced by their charge-conjugate left-handed
fields. The anomaly condition  
\begin{equation}
\mathcal{A}_{ABC}=0  \tag{4.37}  \label{eq:old_4_33}
\end{equation}%
is fulfilled for all gauged generators $\mathbf{X}_{A},\mathbf{X}_{B},%
\mathbf{X}_{C}$.

In the present case the generators $\mathbf{X}_{A}$ include the generators
of color, the generators of weak isospin $\mathbf{J}_{1},\mathbf{J}_{2},%
\mathbf{J}_{3}$ for the group $SU(2)$ , and the tangent-bundle generators $%
\mathbf{T}_{1},\mathbf{T}_{2},\mathbf{T}_{3},\mathbf{E}.$ Hence the complete
anomaly problem is not merely the set of Standard-Model equations involving $%
U(1)_{Y}$. In principle one has to consider the anomalies arising by $%
[SU(3)]^{2}E^{c}(2),[SU(2)]^{2}E^{c}(2),[E^{c}(2)]^{3}$ and $[\mathrm{grav}%
]^{2}E^{c}(2).$

More explicitly, for $\mathbf{X}_{A}\in \{\mathbf{T}_{1},\mathbf{T}_{2},%
\mathbf{T}_{3},\mathbf{E}\},$ the following anomaly coefficients have to
vanish: 
\begin{align}
\mathcal{A}_{33,A}& =\sum_{f}T_{\mathrm{color}}(R_{f})\,\func{tr}_{\mathrm{%
int}}(\mathbf{X}_{A})_{f}=0,  \tag{4.38}  \label{eq:A33A} \\
\mathcal{A}_{22,A}& =\sum_{f}T_{\mathrm{weak}}(R_{f})\,\func{tr}_{\mathrm{int%
}}(\mathbf{X}_{A})_{f}=0,  \tag{4.39}  \label{eq:A22A} \\
\mathcal{A}_{ABC}& =\sum_{f}\func{tr}_{\mathrm{int}}\!\left( \{\mathbf{X}%
_{A},\mathbf{X}_{B}\}\mathbf{X}_{C}\right) _{f}=0,  \tag{4.40}
\label{eq:AABC} \\
\mathcal{A}_{\mathrm{grav},A}& =\sum_{f}\func{tr}_{\mathrm{int}}(\mathbf{X}%
_{A})_{f}=0.  \tag{4.41}  \label{eq:AgravA}
\end{align}%
For non-compact or infinite-dimensional internal representations these
traces have to be understood as traces over the physical regulated
multiplet. This is the same point that appears whenever a chiral anomaly is
evaluated in a mode basis.

In a basis in which $\mathbf{T}_{3}$ and $\mathbf{E}$ are diagonal, the
generators $\mathbf{T}_{1}$ and $\mathbf{T}_{2}$ are non-diagonal
ladder-type operators. Equivalently, $\mathbf{T}_{+}$ and $\mathbf{T}_{-}$
change the $\mathbf{T}_{3}$ eigenvalue. Therefore, for a complete internal
multiplet, 
\begin{equation}
\langle f|\mathbf{T}_{1}|f\rangle =0,\langle f|\mathbf{T}_{2}|f\rangle =0. 
\tag{4.42}  \label{eq:old_4_38}
\end{equation}%
Equivalently, 
\begin{equation}
\func{tr}_{\mathrm{int}}(T_{1})=0,\func{tr}_{\mathrm{int}}(T_{2})=0. 
\tag{4.43}  \label{eq:old_4_39}
\end{equation}%
Thus mixed anomalies with two color generators or two weak-isospin
generators and one insertion of $T_{1}$ or $T_{2}$ vanish by the ladder
selection rule, provided the trace is taken over a complete internal
multiplet.

The nonzero diagonal contributions are those with insertions of $T_{3}$ and $%
E$. These give charge sums involving $Y_{f}$ and $\kappa _{f}$.

For example, the mixed color condition with a diagonal $\mathbf{T}_{3}$
insertion is 
\begin{equation}
\lbrack SU(3)]^{2}\mathbf{T}_{3}:\qquad 2Y_{Q}+Y_{u^{c}}+Y_{d^{c}}=0. 
\tag{4.44}  \label{eq:SU3T3}
\end{equation}%
The mixed weak condition is 
\begin{equation}
\lbrack SU(2)]^{2}\mathbf{T}_{3}:\qquad 3Y_{Q}+Y_{L}=0.  \tag{4.45}
\label{eq:SU2T3}
\end{equation}%
The cubic diagonal condition is 
\begin{equation}
\lbrack \mathbf{T}_{3}]^{3}:\qquad
6Y_{Q}^{3}+3Y_{u^{c}}^{3}+3Y_{d^{c}}^{3}+2Y_{L}^{3}+Y_{e^{c}}^{3}=0. 
\tag{4.46}  \label{eq:T3cubed}
\end{equation}%
The mixed gravitational condition is 
\begin{equation}
\lbrack \mathrm{grav}]^{2}\mathbf{T}_{3}:\qquad
6Y_{Q}+3Y_{u^{c}}+3Y_{d^{c}}+2Y_{L}+Y_{e^{c}}=0.  \tag{4.47}
\label{eq:gravT3}
\end{equation}%
These are the usual hypercharge anomaly equations. In the tangent-bundle
formulation they should be interpreted as the anomaly equations for the
diagonal $\mathbf{T}_{3}$ sector of $E^{c}(2)$.

If the central generator $\mathbf{E}$ is coupled to an independent gauge
field, then analogous anomaly conditions involving $\kappa _{f}$ must also
be considered. They are 
\begin{equation}
\lbrack SU(3)]^{2}\mathbf{E}:\qquad 2\kappa _{Q}+\kappa _{u^{c}}+\kappa
_{d^{c}}=0,  \tag{4.48}  \label{eq:SU3E}
\end{equation}%
\begin{equation}
\lbrack SU(2)]^{2}\mathbf{E}:\qquad 3\kappa _{Q}+\kappa _{L}=0,  \tag{4.49}
\label{eq:SU2E}
\end{equation}%
\begin{equation}
\mathbf{E}^{3}:\qquad 6\kappa _{Q}^{3}+3\kappa _{u^{c}}^{3}+3\kappa
_{d^{c}}^{3}+2\kappa _{L}^{3}+\kappa _{e^{c}}^{3}=0,  \tag{4.50}
\label{eq:E3}
\end{equation}%
and 
\begin{equation}
\lbrack \mathrm{grav}]^{2}\mathbf{E}:\qquad 6\kappa _{Q}+3\kappa
_{u^{c}}+3\kappa _{d^{c}}+2\kappa _{L}+\kappa _{e^{c}}=0.  \tag{4.51}
\label{eq:gravE}
\end{equation}

There can also be mixed diagonal conditions involving both $\mathbf{T}_{3}$
and $\mathbf{E}$. For example, 
\begin{equation}
\mathbf{T}_{3}^{2}\mathbf{E}:\qquad \sum_{f}d_{f}\,Y_{f}^{2}\kappa _{f}=0, 
\tag{4.52}  \label{eq:T3T3E}
\end{equation}%
and 
\begin{equation}
\mathbf{T}_{3}\mathbf{E}^{2}:\qquad \sum_{f}d_{f}\,Y_{f}\kappa _{f}^{2}=0. 
\tag{4.53}  \label{eq:T3EE}
\end{equation}%
Here $d_{f}$ denotes the color and weak-isospin multiplicity of the state $f$%
.

The appearance of $Y_{f}$ in (\ref{eq:SU3T3})--(\ref{eq:gravT3}) refers to
the shifted generator $\mathbf{T}_{3}$ of (\ref{eq:shift_generator}). We can
now derive a central-charge selection under explicit assumptions: one SM
family without a right-handed neutrino is required to cancel its diagonal
anomalies by itself; each weak or color multiplet has one central charge;
and no additional chiral state contributes to these sums. Write all fields
as left-handed Weyl fields and set $a=\kappa _{Q}$, $b=\kappa _{u^{c}}$, $%
c=\kappa _{d^{c}}$, $d=\kappa _{L}$, $k=\kappa _{e^{c}}$. The linear
conditions (\ref{eq:SU3E}), (\ref{eq:SU2E}) and (\ref{eq:gravE}) give 
\begin{equation}
c=-2a-b,\qquad d=-3a,\qquad k=6a.  \tag{4.54}  \label{eq:kappa_linear}
\end{equation}%
Using the hypercharges already fixed in (\ref{eq:observable_assignments}),
the mixed $\mathbf{T}_{3}^{2}\mathbf{E}$ condition (\ref{eq:T3T3E}) becomes 
\begin{equation}
\frac{2}{3}a+\frac{16}{3}b+\frac{4}{3}c+2d+4k=16a+4b=0.  \tag{4.55}
\label{eq:kappa_mixed}
\end{equation}%
Thus $b=-4a$, $c=2a$, and the result is 
\begin{equation}
\kappa _{f}=3aY_{f}.  \tag{4.56}  \label{eq:kappa_proportional}
\end{equation}%
The remaining cubic and mixed diagonal equations then vanish with the SM
hypercharge sums. In right-handed particle notation the charge vector is 
\begin{equation}
(\kappa _{Q},\kappa _{u_{R}},\kappa _{d_{R}},\kappa _{L},\kappa
_{e_{R}})=a(1,4,-2,-3,-6).  \tag{4.57}  \label{eq:kappa_assignment}
\end{equation}%
Since $\kappa _{Q}$ is integer, $a$ is integer. If these observable fields
are also assumed to realize the central $U(1)$ with primitive charges,
meaning that their nonzero integer charges have no common divisor greater
than one, then $|a|=1$. Choosing the central orientation so that $\kappa
_{Q}>0$ gives $a=1$ and hence $\kappa _{f}=3Y_{f}$. This last choice is an
additional normalization and spectrum assumption; anomaly cancellation alone
does not exclude integer multiples or other solutions involving extra fields
or cancellations between families. A primitive central character by itself
does not relate $\mathbf{E}$ and $\mathbf{T}_{3}$. The relation applies to
the selected observable states, not every state in a vertical ladder,
because $\mathbf{T}_{\pm }$ changes $Y$ and preserves $\kappa $.

\bigskip This distinction is also relevant for possible dark-sector states.
A state with 
\begin{equation}
I=0,\qquad Y=0,\qquad \kappa \neq 0  \tag{4.58}  \label{eq:old_4_52}
\end{equation}%
does not contribute to the observable hypercharge anomaly. For dark color
and weak singlets written as left-handed Weyl fields, the additional
conditions from (\ref{eq:E3}) and (\ref{eq:gravE}) reduce to 
\begin{equation}
\sum_a\kappa_a^3=0,\qquad \sum_a\kappa_a=0.  \tag{4.59}  \label{eq:old_4_53}
\end{equation}
A chiral example copies the chirally weighted SM hypercharge pattern: 
\begin{equation}
\begin{array}{c|rrrrr}
\kappa_a & 1 & -4 & 2 & -3 & 6 \\ \hline
\text{number of Weyl fields} & 6 & 3 & 3 & 2 & 1%
\end{array}%
.  \tag{4.60}  \label{eq:old_4_54}
\end{equation}
The sums are $6-12+6-6+6=0$ and $6-192+24-54+216=0$. These numbers count
independent dark species, not color or weak multiplicities. Chirality alone
does not ensure anomaly cancellation. Alternatively a vector-like spectrum
cancels these perturbative gauge anomalies pairwise. This establishes
diagonal central anomaly cancellation for the proposed selected spectrum;
the full vertical transitions must be considered separately.

The remaining possible anomaly structures contain the non-diagonal
generators $\mathbf{T}_{1},\mathbf{T}_{2}$, or equivalently $\mathbf{T}_{+}$
and $\mathbf{T}_{-}$. These terms do not reduce to simple charge sums. Their
traces have the schematic form 
\begin{equation}
\func{tr}_{\mathrm{int}}\!\left( \{\mathbf{T}_{+},\mathbf{T}_{-}\}\mathbf{T}%
_{3}\right) ,  \tag{4.61}  \label{eq:TpTmT3}
\end{equation}%
\begin{equation}
\func{tr}_{\mathrm{int}}\!\left( \{\mathbf{T}_{+},\mathbf{T}_{-}\}\mathbf{E}%
\right) ,  \tag{4.62}  \label{eq:TpTmE}
\end{equation}%
or similar expressions. Such terms are absent in a purely diagonal $U(1)$
theory, but they may occur in the full $E^{c}(2)$ gauge theory.

The non-diagonal generators $\mathbf{T}_{\pm }=\mathbf{T}_{1}\pm i\mathbf{T}%
_{2}$ define transition currents. Before expansion into four-dimensional
modes, the current is%
\begin{equation}
j_{B^{\pm }}^{\mu }=\int d\mathbf{\xi }\overline{\psi }(x,\mathbf{\xi }%
)\gamma ^{\mu }\mathbf{T}_{\pm }\psi (x,\mathbf{\xi })  \tag{4.63}
\label{eq:old_4_57}
\end{equation}

Expanding the tangent-bundle field in internal eigenfunctions gives the
current 
\begin{equation}
J_{B^{\pm }}^{\mu }=\sum_{f,g}\overline{\psi _{f}}(x)\gamma ^{\mu }\psi
_{g}(x)\prec f\mid \mathbf{T}_{\pm }\mid g\succ  \tag{4.64}
\label{eq:old_4_58}
\end{equation}

where%
\begin{equation}
\prec f\mid T_{\pm }\mid g\succ =\int d\mathbf{\xi }h_{f}^{\ast }(\mathbf{%
\xi })\mathbf{T}_{\pm }h_{g}(\mathbf{\xi })  \tag{4.65}  \label{eq:old_4_59}
\end{equation}

Thus $J_{B^{\pm }}^{\mu }$ is a transition current between different
internal states. The interaction with a $B_{\pm }$-boson exists only if the
transition matrix element $\prec f\mid \mathbf{T}_{\pm }\mid g\succ $ is
nonzero.

If all transition matrix elements in (\ref{eq:old_4_59}) vanish between
retained states, the projected current in (\ref{eq:old_4_58}) is identically
zero. The corresponding truncated fermion theory has no direct $B^{\pm }$
vertices and hence no fermion triangle diagrams with such external vertices.
This is the relevant statement for the projected lepton analysis of Ref.~%
\cite{I1}; the same criterion can be checked for the selected quark modes.

The selected SM family has four left-handed weak doublets (three colored
quark copies and one lepton doublet), satisfying the global $SU(2)$
doublet-parity condition \cite{WittenAnomaly}.

\textbf{4.6 Renormalization and the one-loop beta function}

The discovery of asymptotic freedom of non-Abelian gauge theories gave a
solid theoretical foundation of QCD turned it from a speculative idea into a
predictive quantum field theory. The beta function describes how the
coupling changes with energy scale. In QCD, the one-loop calculation gave a
negative beta function, meaning the coupling runs downward as the energy
increases; that sign is the mathematical origin of asymptotic freedom \cite%
{Gross,Pol} .

In this section we show on a Lagrangian level that renormalization constants
and beta function in the emergent QCD are identical to those of ordinary QCD
under the stated projection assumptions.

Using traditional notations for QCD renormalization constants $Z_{i}.$the
rescaling can be written as

\begin{equation}
G_{0\mu }^{A}=Z_{3}^{1/2}G_{\mu
}^{A},q_{0}=Z_{2}^{1/2}q,c_{0}^{A}=Z_{c}^{1/2}c^{A},g_{0}=\mu ^{\varepsilon
}Z_{g}g.  \tag{4.66}  \label{eq:old_4_60}
\end{equation}

where $c^{A}$ denotes the Faddeev-Popov ghost field in $d=4-2\epsilon .$ By
comparison of the Lagrangian for the bare fields with the renormalzed
quark-gluon Lagrangian we easy find a relation for the renormalized coupling
constant 
\begin{equation}
Z_{g}=\frac{Z_{1}}{Z_{2}Z_{3}^{1/2}},\qquad g_{0}=\mu ^{\varepsilon }Z_{g}g 
\tag{4.67}  \label{eq:old_4_61}
\end{equation}

Renormalization of the three-gluon vertex and the ghost-gluon vertex implies
additional relations among the renormalization constants $Z_{i}.$ As
guaranteed by BRST invariance and Slavnov-Taylor identities, all of these
separate relations are compatible and yield the same $\beta $ function.The
quark field is expanded in the three degenerate internal modes,%
\begin{equation}
q(x,\xi )=\sum_{c=1}^{3}q_{c}(x)h_{c}(\mathbf{\xi }),  \tag{4.68}
\label{eq:old_4_62}
\end{equation}%
The internal functions are assumed to be orthonormal and 
\begin{equation}
\func{tr}(T^{A}T^{B})=T(F)\delta ^{AB},\qquad C_{A}=3,C_{F}=\frac{4}{3},T(F)=%
\frac{1}{2},  \tag{4.69}  \label{eq:old_4_63}
\end{equation}%
without extra vertical form factors in the degenerate-triplet approximation.
BRST symmetry follows from gauge invariance of the projected action, not
from the loop integrations. Gauge fixing and the Faddeev--Popov construction
give the ordinary nilpotent $SU(3)$ BRST transformations, whose invariance
yields the Slavnov--Taylor identities. The functions $h_{i}$ enter only
through the projection and do not modify these identities.

The Slavnov--Taylor identities ensure that all gauge vertices give the same $%
Z_{g}$. In minimal subtraction, 
\begin{equation}
Z_{g}=1-\frac{g^{2}}{16\pi ^{2}\varepsilon }\left[ \frac{11}{6}C_{A}-\frac{2%
}{3}T(F)N_{f}\right] +O(g^{4}),  \tag{4.70}  \label{eq:Zg}
\end{equation}%
where $N_{f}$ is the number of active Dirac quark flavours. The counterterm
is of order $g^{2}$.

Since $g_{0}$ is independent of $\mu $, 
\begin{equation}
0=\mu \frac{dg_{0}}{d\mu }=\mu ^{\varepsilon }Z_{g}\left[ \varepsilon
g+\beta (g)+g\mu \frac{d}{d\mu }\ln Z_{g}\right] .  \tag{4.71}
\label{eq:old_4_65}
\end{equation}%
and 
\begin{equation}
\mu \frac{d}{d\mu }\ln Z_{g}=\frac{g^{2}}{16\pi ^{2}}\left[ \frac{11}{3}%
C_{A}-\frac{4}{3}T(F)N_{f}\right] +O(g^{4}).  \tag{4.72}  \label{eq:dlnZg}
\end{equation}%
Consequently, the renormalization-group equation for the coupling has the
form 
\begin{align}
\beta (g)& =-\varepsilon g-g\mu \frac{d}{d\mu }\ln Z_{g}=  \tag{4.73}
\label{eq:beta-intermediate} \\
& =-\varepsilon g-\frac{g^{3}}{16\pi ^{2}}\left[ \frac{11}{3}C_{A}-\frac{4}{3%
}T(F)N_{f}\right] +O(g^{5}).  \notag
\end{align}%
Substituting into equ.(\ref{eq:beta-intermediate}) C$_{A}=3$ and T$(F)=\frac{%
1}{2}$ and taking $\varepsilon \rightarrow 0$ gives 
\begin{equation}
\beta (g)=-\frac{g^{3}}{16\pi ^{2}}(11-\frac{2}{3}N_{f})+O(g_{s}^{5}) 
\tag{4.74}  \label{eq:old_4_68}
\end{equation}%
The color sector is asymptotically free for $N_{f}<33/2$. A Standard-Model
family contains two quark flavours, and only active flavours are counted.
This agrees with the SM result \cite{Gross,Pol} and establishes one-loop
perturbative universality if the effective theory is local, the triplet is
exactly degenerate, and higher vertical levels do not mix with it.

\textbf{5. Relation to teleparallel gravity and its extension}

Teleparallel gravity is a reformulation of Einstein's general relativity in
which gravitation is described by torsion rather than by the curvature of
the gravitational connection \cite{AL,t1,t2,t3}. The metric and the
predictions of the theory of general relativity are retained, but the
geometric variables and their interpretation are different. A coframe, or
tetrad, relates coordinate components to components in a local Lorentz
frame. Together with a flat, metric-compatible spin connection, it defines a
Weitzenb\"{o}ck geometry. This connection has vanishing curvature but can
have nonzero torsion. The Levi--Civita connection of the same metric, used
in general relativity, instead has zero torsion and generally nonzero
curvature. The equivalence concerns their gravitational field equations; it
does not identify the two connections.

The translational part of the affine tangent-fiber structure discussed in
Section 2 supplies the natural connection with this description of gravity.
Its translational potential $\theta ^{a\text{ }}$is identified with the
coframe $e^{a}$. The gravitational spin connection $\omega _{\mu }^{ab}$
accounts for inertial effects and can locally be set to zero in the Weitzenb%
\"{o}ck gauge \cite{KP}. In the present framework an additional
Lorentz-sector potential may represent non-gravitational internal fields.
The distinction between that potential and the flat gravitational spin
connection is essential. We first state the ordinary teleparallel theory and
then describe a possible interaction with this additional sector at the
level of the action and field equations.

5.1 Translational geometry and the gravitational action

Let $\mathbf{P}_{a}$ denote a translation generator and $\mathbf{J}_{ab}$ a
Lorentz generator. The coframe and the corresponding affine connection are 
\begin{equation}
\theta ^{a}=e^{a}{}_{\mu }dx^{\mu },\qquad \mathcal{A}_{\mathrm{TP}}=\theta
^{a}\mathbf{P}_{a}+\frac{1}{2}\omega _{\mathrm{TP}}^{ab}\mathbf{J}_{ab}. 
\tag{5.1}  \label{eq:affine_gravity}
\end{equation}%
The dual frame is $e_{a}=e_{a}{}^{\mu }\partial _{\mu }$, with $e_{a}{}^{\mu
}e^{b}{}_{\mu }=\delta _{a}{}^{b}$. Thus $\theta ^{a}=e^{a}$ is a one-form,
whereas $e_{a}$ is a vector. The metric and volume density are $g_{\mu \nu
}=\eta _{ab}e^{a}{}_{\mu }e^{b}{}_{\nu }$ and $e=\det (e^{a}{}_{\mu })$,
with an oriented coframe and $\eta _{ab}=\mathrm{diag}(-1,1,1,1)$. The
background torsion is 
\begin{equation}
R^{a}{}_{b}[\omega _{\mathrm{TP}}]=0,\qquad T_{\mathrm{TP}}^{a}=d\theta
^{a}+\omega ^{a}{}_{b\mathrm{TP}}\wedge \theta ^{b}.  \tag{5.2}
\label{eq:TPbackground}
\end{equation}%
$T_{\mathrm{TP}}^{a}$ is the field strength associated with the geometry of
standard teleparallel gravity.

For any tensor $U^\rho{}_{\mu\nu}$ antisymmetric in its last two indices,
define the quadratic torsion scalar by 
\begin{equation}
\mathcal{T}(U)=\frac14U^\rho{}_{\mu\nu}U_\rho{}^{\mu\nu} +\frac12U^\rho{}%
_{\mu\nu}U^{\nu\mu}{}_{\rho}-u_\mu u^\mu, \qquad u_\mu=U^\rho{}_{\mu\rho}. 
\tag{5.3}  \label{eq:Tscalar}
\end{equation}
Spacetime indices are raised with $g_{\mu\nu}$. The relative sign of the
action is fixed by its relation to general relativity. With the Levi--Civita
Ricci scalar $\mathring R$ and the curvature convention $\mathring R^\rho{}%
_{\sigma\mu\nu} =\partial_\mu\mathring\Gamma^\rho{}_{\nu\sigma}
-\partial_\nu\mathring\Gamma^\rho{}_{\mu\sigma}+\cdots$, the identity is 
\begin{equation}
\mathring R=-\mathcal{T}(T_{\mathrm{TP}}) +\frac2e\partial_\mu(e
v^\mu),\qquad v^\mu=T_{\mathrm{TP}}^\nu{}_{\nu}{}^\mu.  \tag{5.4}
\label{eq:TEGR_identity}
\end{equation}
Consequently the action 
\begin{equation}
S_{\mathrm{TP}}=-\frac1{2\kappa_G} \int d^4x\,e\,\mathcal{T}(T_{\mathrm{TP}%
}),\qquad \kappa_G=8\pi G,  \tag{5.5}  \label{eq:TP_action}
\end{equation}
differs from the Einstein--Hilbert action $\int e\mathring R/(2\kappa_G)$
only by a boundary term. Variations with appropriate boundary conditions
therefore give Einstein's equation. The minus sign in (\ref{eq:TP_action})
follows from the scalar and metric conventions just specified; it is
reversed if the torsion scalar is defined with the opposite overall sign.

For later use, define the gravitational Euler derivative and the source
tensor of an action $S_s$ by 
\begin{equation}
\mathcal{E}_a{}^\nu=-\frac{\kappa_G}{e} \frac{\delta S_{\mathrm{TP}}}{\delta
e^a{}_{\nu}},\qquad \Theta_a{}^\nu{}_{s}=\frac1e\frac{\delta S_s}{\delta
e^a{}_{\nu}}.  \tag{5.6}  \label{eq:source_definitions}
\end{equation}
For the ordinary teleparallel theory, $\mathcal{E}_a{}^\nu=e_a{}^\mu%
\mathring G_\mu{}^\nu$, where $\mathring G_\mu{}^\nu$ is the Einstein tensor
of the metric. These variational definitions fix the normalization and sign
of the field equations without requiring a separate component convention for
the teleparallel superpotential.

5.2 Additional TF Lorentz-sector potential and the interaction action

An additional Lorentz-valued one-form $\Omega ^{a}{}_{b}$ arising from the
non-gravitational gauge fields gives 
\begin{equation}
\widehat{\omega }^{a}{}_{b}=\omega ^{a}{}_{b\mathrm{TP}}+\Omega
^{a}{}_{b},\qquad \widehat{T}^{a}=T_{\mathrm{TP}}^{a}+X^{a},\qquad
X^{a}=\Omega ^{a}{}_{b}\wedge \theta ^{b}.  \tag{5.7}
\label{eq:general_torsion_addition}
\end{equation}%
In components this reads 
\begin{equation}
X^{a}{}_{\mu \nu }=\Omega _{\mu }{}^{a}{}_{b}e^{b}{}_{\nu }-\Omega _{\nu
}{}^{a}{}_{b}e^{b}{}_{\mu },\qquad X^{\rho }{}_{\mu \nu }=e_{a}{}^{\rho
}X^{a}{}_{\mu \nu }.  \tag{5.8}  \label{eq:Xcomponents}
\end{equation}%
The connection $\widehat{\omega }$ is generally not teleparallel, since 
\begin{equation}
R[\widehat{\omega }]=D_{\mathrm{TP}}\Omega +\Omega \wedge \Omega .  \tag{5.9}
\label{eq:extended_curvature}
\end{equation}%
Here $D_{\mathrm{TP}}$ is the TF Lorentz-covariant exterior derivative
formed with the flat background connection. Thus the construction is an
extension on a teleparallel background. The flat connection and its ordinary
teleparallel gravitational action remain separately defined.

Only an internal generator with a specified action on tangent-vector indices
can contribute to $\Omega ^{a}{}_{b}$ in (\ref{eq:general_torsion_addition}%
). A weak or color generator acting on a multiplicity space does not acquire
such an action merely from the dimension of that space. Likewise, the
central potential $C_{\mu }$ has no direct contribution to $X^{a}$ when the
central generator acts trivially on tangent vectors. These fields retain
their usual matter and gauge couplings. We leave the projection onto
tangent-vector generators unspecified here and describe the interaction in
terms of the resulting tensor $X^{a}$.

Denote the matter and non-gravitational gauge actions of the preceding
sections by $S_{\mathrm{m}}$ and $S_{\mathrm{gauge}}$. One possible extended
action is 
\begin{align}
S&=S_{\mathrm{TP}}+S_{\mathrm{m}}+S_{\mathrm{gauge}}+S_{\mathrm{int}}, 
\notag \\
S_{\mathrm{int}}&=-\frac1{2\kappa_G}\int d^4x\,e\, \left[\mathcal{T}(T_{%
\mathrm{TP}}+\lambda X) -\mathcal{T}(T_{\mathrm{TP}})\right].  \tag{5.10}
\label{eq:Sint}
\end{align}
The parameter $\lambda$ is dimensionless when $X$ and $T_{\mathrm{TP}}$ have
the same dimensions. It measures an additional interaction; its value is not
fixed by the torsion identity. Because $\mathcal{T}$ is quadratic, (\ref%
{eq:Sint}) is exactly 
\begin{equation}
S_{\mathrm{int}}=-\frac1{2\kappa_G}\int d^4x\,e\, \left[2\lambda\mathcal{C}%
(T_{\mathrm{TP}},X)+ \lambda^2\mathcal{Q}(X)\right].  \tag{5.11}
\label{eq:CQ_action}
\end{equation}
Writing $T=T_{\mathrm{TP}}$, $t_\mu=T^\rho{}_{\mu\rho}$ and $x_\mu=X^\rho{}%
_{\mu\rho}$, the two scalar contractions are 
\begin{align}
\mathcal{C}(T,X)&=\frac14T^\rho{}_{\mu\nu}X_\rho{}^{\mu\nu} +\frac14\left(T^%
\rho{}_{\mu\nu}X^{\nu\mu}{}_{\rho} +X^\rho{}_{\mu\nu}T^{\nu\mu}{}_{\rho}%
\right)-t_\mu x^\mu,  \notag \\
\mathcal{Q}(X)&=\frac14X^\rho{}_{\mu\nu}X_\rho{}^{\mu\nu} +\frac12X^\rho{}%
_{\mu\nu}X^{\nu\mu}{}_{\rho}-x_\mu x^\mu.  \tag{5.12}  \label{eq:CQ_general}
\end{align}
The first term mixes gravitational torsion with the additional torsion; the
second contains the contribution quadratic in $X$. Both are defined without
choosing a particular gauge field or a cosmological background.

This action specifies a structural coupling ansatz. Its use for a given
internal gauge sector requires a compatible tangent-vector projection and
transformation laws.

5.3 Field equations and scope of the extension

Varying the coframe gives 
\begin{equation}
\mathcal{E}_a{}^\nu=\kappa_G \left(\Theta_a{}^\nu{}_{\mathrm{m}} +\Theta_a{}^%
\nu{}_{\mathrm{gauge}} +\Theta_a{}^\nu{}_{\mathrm{int}}\right),\qquad
\Theta_a{}^\nu{}_{\mathrm{int}} =\frac1e\frac{\delta S_{\mathrm{int}}}{%
\delta e^a{}_{\nu}}.  \tag{5.13}  \label{eq:extended-field-equation}
\end{equation}
The ordinary energy-momentum of the gauge fields is already included in $%
\Theta_{\mathrm{gauge}}$. The extra tensor $\Theta_{\mathrm{int}}$ describes
the additional backreaction arising from the proposed torsion coupling. Its
variation includes the dependence of $X^a$ on the coframe, as well as the
determinant and all metric contractions. It must be taken before imposing a
symmetry-restricted gravitational ansatz. The ordinary teleparallel field
equations are recovered for $\lambda=0$.

The same interaction also modifies the equation for any gauge potential
entering the chosen projection. With the gauge form $\widetilde h_{AB}$ of
Section 2.4, its structural form is 
\begin{equation}
\frac1eD_\mu\left(e\widetilde h_{AB}F^{B\mu\nu}\right) =J_{A\mathrm{m}%
}{}^\nu+J_{A\mathrm{int}}{}^\nu,\qquad J_{A\mathrm{int}}{}^\nu=-\frac1e 
\frac{\delta S_{\mathrm{int}}}{\delta A_\nu^A}.  \tag{5.14}
\label{eq:general_gauge_current}
\end{equation}
Here $D_\mu$ is the corresponding internal gauge-covariant derivative.
Equations (\ref{eq:extended-field-equation}) and (\ref%
{eq:general_gauge_current}) describe the two directions of the coupling: the
internal fields contribute to the gravitational equation, and the torsion
interaction contributes to their gauge equations.

The translational structure therefore provides the ordinary teleparallel
sector and a systematic way to formulate additional couplings to TF
Lorentz-sector fields. The present discussion establishes their general
action and variational structure. Identifying a physical projection, fixing
its coupling and solving the resulting field equations are separate
dynamical tasks. Such applications are left for future work.

\textbf{6. Conclusion and outlook}

In this work, a quantum-field-theoretical framework on the tangent bundle
has been developed in which gravitational and internal gauge structures are
associated with different sectors of the tangent-fiber geometry. The
construction uses an affine frame structure whose Lie algebra is isomorphic
to the Poincar\'{e} algebra. This algebra acts on affine tangent frames over
a fixed spacetime point. In the interpretation adopted here, selected
components of the Lorentz-sector connection are associated with electroweak
and additional internal gauge potentials, while the translational sector
supplies the gravitational coframe and its relation to teleparallel gravity.

The reduced vertical spectral problem is based on the Euclidean group $E(2)$
and its central extension $E^{c}(2)$. Equivalent copies of vertical $E^{c}(2)
$ representations admit unitary symmetries acting on their multiplicity
spaces. For the twofold multiplicity of the selected left-chiral sectors and
the assumed threefold quark multiplicity, these include the weak $SU(2)_{L}$
and color $SU(3)_{c}$ actions, respectively. The construction thus relates
these internal symmetries to the representation structure of the
tangent-fiber Hilbert space. The local internal structure $SU(3)_{c}\times
SU(2)_{L}\times E^{c}(2)$ is not a subgroup decomposition of $SO(3,1)$: the $%
E^{c}(2)$ factor acts on the irreducible vertical representation space,
whereas the weak and color factors act on equivalent-copy spaces. The
proposed global identification correlates color-center phases with vertical
rotations and, together with the selected chiral matter spectrum, anomaly
conditions and charge normalization, reproduces the observed quark
hypercharges.

The quantum-field-theoretic consequences of this construction have been
formulated for the selected particle sectors. Matter fields are expanded in
internal eigenfunctions $h_{nm\varkappa}(\boldsymbol{\xi})$ of the vertical $%
E^{c}(2)$ Casimir operator. Their expansion coefficients are spacetime
quantum fields, while the internal quantum numbers label the vertical
states. Orthonormality of the internal basis preserves canonical field
quantization after projection. Integration over the vertical coordinates
yields spacetime interactions whose couplings retain the appropriate
internal matrix elements. In the color sector, the generators act on
equivalent copies and leave the normalized vertical wavefunction unchanged.
Under the stated projection and locality assumptions, the resulting action
has the standard local QCD structure, including its BRST symmetry, one-loop
beta function and asymptotic freedom. Gauge fields are treated as spacetime
connections and are not expanded in the matter Casimir eigenfunctions.

Beyond the SM gauge fields, the construction contains hypothetical $B^{\pm }$
bosons associated with $\mathbf{T}_{1},\mathbf{T}_{2}$ and an additional
neutral potential $C_{\mu }$ associated with the central generator $\mathbf{E%
}$. Neutral mixing determines the physical combinations in the assumed
symmetry-breaking pattern. States with $I=Y=0$ and nonzero central charge $%
\varkappa $ provide hypothetical dark-matter candidates. The diagonal
anomaly conditions admit the explicit chiral dark spectrum given in
Section~4.5. Under the single-family and spectrum assumptions stated there,
the observable central charges are proportional to hypercharge; a primitive
integer choice fixes their minimal common normalization.

The translational geometry has been related to the teleparallel equivalent
of general relativity. A proposed extension is formulated through an
additional Lorentz-valued potential and the mixed and quadratic contractions
of its induced torsion. The action yields an additional source in the
gravitational equation and a corresponding contribution to the gauge-field
equations. This provides a structural framework for studying such couplings;
their physical projections, strengths and dynamical consequences require
further investigation.

A distinctive direction for future work is the internal structure of the
vacuum. In the proposed interpretation, the $n=0$ spectral sector is
distinguished from the higher levels assigned to observable particles. The
dependence of its wavefunctions on continuous internal coordinates,
including $\xi _{1},\xi _{2}$, allows an internal structure whose dynamics
remain to be investigated. Possible condensates and their interactions with
observable particle sectors may provide a mechanism for mass generation and
family-dependent scales. These are proposed dynamical possibilities rather
than mass predictions established here. The corrected Casimir spectrum
provides the spectral basis for such investigations.

\textbf{Data availability}

No datasets were generated or analysed in this theoretical study.


\begin{thebibliography}{99}
\raggedright

\bibitem{Lu} E. Lubkin, Geometric definition of gauge invariance, Annals of
Physics 23, 233--283 (1963).

\bibitem{Trau} A. Trautman, "Fiber bundles, gauge fields and gravitation"
Rep. Math. Phys. 1(1), 29 (1970),

\bibitem{Wu} T.T. Wu and C.N. Yang, "Concept of nonintegrable phase factors
and global formulation of gauge fields", Phys. Rev. D 12, 3845 (1975),

\bibitem{Ati} M.F. Atiyah "The geometry of Yang-Mills fields", Scuola
Normale Superiore, Pisa, Italy (1979),

\bibitem{Nak} M. Nakahara, "Geometry, Topology and Physics", Institute of
Physics , Bristol (1990),

\bibitem{Is} C. J. Isham,\textquotedblright Modern Differential Geometry for
Physicists\textquotedblright , World Scientific (1999),

\bibitem{I1} J. Herrmann,"Extension of the standard model of electroweak
interaction and Dark Matter in the tangent bundle geometry", Eur. Phys. J. C
79, 779 (2019), ArXiv:1802.03228v3

\bibitem{I2} J. Herrmann,"Towards a unified theory of the fundamental
physical interactions based on the underlying geometrical structure of the
tangent bundle" Eur. Phys. J. C 82,947 (2022), ArXiv:2207.14557v1

\bibitem{I3} J. Herrmann, "Origin of internal symmetries of the fundamental
interactions,the family problem, fractional quark charges, and unification
in the tangent bundle geometry", LHEP-427,2023,.

\bibitem{AL} R. Aldrovandi, J. G. Pereira, Teleparallel Gravity: An
Introduction (Springer, Dordrecht, 2013).

\bibitem{P1} T. W. Kibble, J. Math. Phys. 2, 212, (1961),

\bibitem{P2} D. W. Sciama, in "Recent Developements in General Relativity"
(Pergamon, New York, 1962) p.415

\bibitem{P3} F.W. Hehl, J.D. McCrea, E.W. Mielke and Y. Ne\'{}eman, Phys.
Rep. 258, 1 (1995)

\bibitem{t1} K. Hayashi and T. Nakno, Prog Theor. Phys. 38, 491 (1967),

\bibitem{t2} Y. M. Cho, Phys. Rev. D14, 2521 (1976),

\bibitem{t3} K. Hayashi and T. Shirafuji, Phys. Rev. D19, 3524 (1979),

\bibitem{t4} J. W. Maluf, Annalen Phys. 525, 339 (2013),arXiv:1303.3897
[gr-qc]

\bibitem{g5} T. Dass, Pramana \textbf{23}, 433 (1984),

\bibitem{Hall} B. C. Hall, Lie Groups, Lie Algebras, and Representations: An
Elementary Introduction, 2nd edn. (Springer, Cham, 2015).

\bibitem{MS1} A. A. Mele, Lecture Notes on Representation Theory for Quantum
Information, QMATH Masterclass 2025, Copenhagen

\bibitem{MS2} S. Moudgalya and O. I. Motrunich, "Hilbert Space Fragmentation
and Commutant Algebras", Phys. Rev. X 12, 011050 (2022)

\bibitem{r10} H. Hoogland, J. Phys. A Math. Gen. 11, 1557 (1978),

\bibitem{r11} J. F. Cari\~{n}ena, M. A. del Olmo and M Santander, J. Phys.
A: Math. Gen. 17, 3091 (1984).

\bibitem{LiWang} C.-F. Li and Q. Wang, The quantum behavior of an electron
in a uniform magnetic field, Physica B 269, 22--27 (1999).
https://doi.org/10.1016/S0921-4526(99)00045-9.

\bibitem{Quinn} J. J. Quinn, A. W\'{o}js, K.-S. Yi and G. Simion, The
hierarchy of incompressible fractional quantum Hall states, arXiv:0906.5182
[cond-mat.str-el] (2009). https://doi.org/10.1016/j.physrep.2009.06.002.

\bibitem{Na} C. R. Nappi and E. Witten, \textquotedblleft A WZW model based
on a non-semi-simple group,\textquotedblright\ Phys. Rev. Lett. 71,
3751--3753 (1993), arXiv:hep-th/9310112.

\bibitem{Tsey} A. A. Tseytlin, \textquotedblleft On gauge theories for
non-semisimple groups,\textquotedblright\ Nucl. Phys. B 450, 231--250
(1995), arXiv:hep-th/9505129.

\bibitem{H2} S.C. Zhang, T. H. Hansson and S.Kivelson, Phys. Rev. Lett.62,
82, (1989),

\bibitem{H3} A. Lopez and E. Fradkin, Phys. Rev. B 44, 5246 (1991),

\bibitem{Bery} M.~V.~Berry, Quantal phase factors accompanying adiabatic
changes, Proc.\ Roy.\ Soc.\ A \textbf{392} (1984) 45.

\bibitem{Wil} F.Wilczek and A. Zee, "Appearance of Gauge Structure in Simple
Dynamical Systems". Physical Review Letters, 52(24), 2111(1984)

\bibitem{Tong} D. Tong, "Line operators in the Standard Model", JHEP 07,104
(2017)

\bibitem{Aha} O. Aharony, N. Seiberg and Y.Tachikawa,"Reading between the
lines of four-dimensional gauge theories", JHEP 08, 115 (2013).

\bibitem{H.Lin} H.-L. Li and L.-X. Xu, Understanding the SM gauge group from
SMEFT, JHEP 07, 199 (2024), arXiv:2404.04229.
https://doi.org/10.1007/JHEP07(2024)199.

\bibitem{Ban} G. Bandelloni, C. Becchi, A. Blasi, and R. Collina,
\textquotedblleft Non semisimple gauge models. I. Classical theory and the
properties of ghost states,\textquotedblright\ Ann. Inst. H. Poincar\'{e} A
28, 225--254(1978)

\bibitem{Wig} E. P. Wigner, Ann. Math. 40, 149 (1939),

\bibitem{Bar} V. Bargmann, \textquotedblleft On unitary ray representations
of continuous groups,\textquotedblright\ Annals of Mathematics 59, 1--46
(1954).

\bibitem{Wein} S. Weinberg, "The Quantum Theory of Fields", Vol. 2 Modern
Applications (Cambridge University Press, Cambridge, 1996).

\bibitem{Fer} F. Ferrari, \textquotedblleft Chern--Simons field theories
with non-semisimple gauge group of symmetry,\textquotedblright\ J. Math.
Phys. 44, 138--156 (2003), arXiv:hep-th/0210100.

\bibitem{BRST} C. Becchi, A. Rouet, and R. Stora, \textquotedblleft
Renormalization of Gauge Theories,\textquotedblright\ Annals of Physics 98,
287 (1976)

\bibitem{Pes} M.Peskin, D.V. Schroeder, "An Introduction to Quantum Field
Theory", Westview Press 1995

\bibitem{WittenAnomaly} E. Witten, An SU(2) anomaly, Phys. Lett. B 117,
324--328 (1982). https://doi.org/10.1016/0370-2693(82)90728-6.

\bibitem{Gross} D. J. Gross and F. Wilczek, \textquotedblleft Ultraviolet
Behavior of Non-Abelian Gauge Theories,\textquotedblright\ Physical Review
Letters 30, 1343 (1973).

\bibitem{Pol} H. D. Politzer, \textquotedblleft Reliable Perturbative
Results for Strong Interactions?\textquotedblright\ Physical Review Letters
30, 1346 (1973).

\bibitem{KP} M. Kr\v{s}\v{s}\'{a}k and J. G. Pereira, "Spin connection and
renormalization of teleparallel action", Eur. Phys. J. C 75, 519 (2015),
arXiv:1504.07683.
\end{thebibliography}
\end{document}